# Gas Storage Potential of Li-decorated $ExBox_4^+$

***Ranjita Das and Pratim Kumar Chattaraj****

Department of Chemistry and Center for Theoretical Studies, Indian Institute of Technology

Kharagpur, Kharagpur – 721302, West Bengal, India

Correspondence to: Pratim K. Chattaraj (Email: pkc@chem.iitkgp.ernet.in)

**Abstract**: The newly developed compound $ExBox^{4+}$ is explored to check whether it is a proficient hydrogen storage material. Both exoherdal and endohedral hydrogen adsorption on $ExBox^{4+}$ are studied. Endohedral hydrogen molecules interact strongly than exohedral ones. The hydrogen adsorption energy is as good as the recently studied charged fullerenes. The hydrogen storage capacity appears to be ~4.3 wt%. The endohedral CO sorption is also analysed with the help of DFT. The first principle DFT calculation and MD simulation are performed to investigate the effect of lithium doping on the gas adsorbing capacity and adsorption enthalpy as well as adsorption energy of $ExBox^{4+}$.The metal atom interaction with $ExBox^{4+}$is found to be pretty strong, and the interaction energy appears to be higher than the metal cohesive energy.The thermodynamic parameters suggest that metal doping method is spontaneous in nature. The analysis of adsorption energy, thermodynamic properties and MD simulation results suggest that Li doped $ExBox^{4+}$ can be a promising gas adsorbing material. Lithium doping increases the gas adsorption capacity of $ExBox^{4+}$. The Li decorated $ExBox^{4+}$ is found to adsorb twenty four hydrogen molecules with a gravimetric wt% of 6.23. Adsorption of CO on the metal decorated $ExBox^{4+}$ is also studied. AIM analysis is performed to obtain a general idea of the bonding interaction between Li-H and Li-C.

## Introduction:

The adsorption of small gas molecules, like $H_2$, CO, $CO_2$, $N_2$ etc. on the metal decorated nano-structures have been an interesting subject of study for the past several years[1-5]. Adsorption of gas molecules is important due to their industrial usage and biological impotence. Hydrogen ($H_2$) has been accepted as an attractive alternative energy carrier, since it is lightweight, non-polluting, highly efficient, and easily derived. Hence material for efficient and cost-effective storage of hydrogen has been exposed by the scientific community. On the other hand CO is a toxic gas for human and animals, carbon monoxide poisoning causes toxicity of the central nervous system and heart, even causes death. Highest tolerance level of CO is ~ 50 ppm for a time weighted average limit (TWA) in air. Hence exploration of certain gas-sensing materials with high sensitivity and selectivity to CO is highly important. Carbon monoxide adsorbent is also required for long-lasting laboratory usage. Further, for many heterogeneous catalytic reactions CO adsorption is an important mechanistic step in the chemical reaction pathway.

Several scientific studies[6,7]explored that for effective gas adsorption the material should exhibit gas storage capacity and a moderate gas binding energy. An effective gas adsorption requires high accessible surface area as well as large free pore volume.

Some theoretical and experimental studies are involved in finding potential hydrogen storage materials. A potential hydrogen storage material should consist of light weighted elements with wide surface area. Carbon nano materials like carbon nano tube (CNT), fullerene, graphene, are studied for their hydrogen binding ability.The gas adsorption capacity of several materials appear to be influenced either by metal doping or inclusion of charge. It is worth mentioning here that the hydrogen storage capacity of CNT gets increased in presence of adsorbed metal dopants[8]. Since the driving force for $H_2$ adsorption is the van der Waals interaction, the presence of metal atoms increases the adsorption enthalpy by a considerable amount.[9-12]Srinivasu et al explored the applicability of transition metal decorated porphyrin-like fullerene as potential hydrogen storage material.[13]In a first principle density functional calculation Guo et al. reported hydrogen storage capacity of Li decorated graphyne.[14] Recently Bodrenko and coworkers[15] studied some metal cation stabilized (alkali and alkaline earth metal ions) anionic aromatic carbon-based rings for their hydrogen adsorption capacity and found that the properties of the cation is the governing factor for adsorption rather than the ring itself. Further some recent studies[16] reveal the capacity of charged carbon fullerenes $C_n$ ($20 \leq n \leq 82$) as hydrogen storage media.

In a combined experimental and theoretical study[17] the adsorption of CO on CoMo and NiMosulfide catalysts is analysed. A combined quantum chemical and experimental IR spectra study explored the adsorption of CO on hydroxyapatite [HA, $Ca_{10}(PO_4)_6(OH)_2$]. Interestingly they found that their suggested method was able to analyse that the relative contribution of the stoichiometric, *Ca-rich and P-rich surfaces*, to the surface terminations of the layers dictate the morphology of HA nanoparticles. Sun et al[18] recently studied the adsorption of CO on the stoichiometric, Ca-doped, and defective (oxygen vacancy) $LaFeO_3$ (010) surface by means of a first-principles calculations based on density functional theory (DFT)[19-21]. They found that CO acts as a donor in adsorption process when adsorbed on the stoichiometric and Ca-doped $LaFeO_3$ (010) surface with the Fe-CO configuration, but CO act as acceptor when adsorbed on the defective (oxygen vacancy) $LaFeO_3$ (010) surface with the defect-CO configuration. In a volumetric measurement and grand canonical Monte Carlo simulation study[22]adsorption of CO, $CH_4$, and $N_2$ in zeolite-X exchanged with different alkali metal ions was explored. The adsorption of CO and $CH_4$ appeared to show higher sorption capacity than $N_2$ for all cation exchanged zeolite samples.

Recently Stoddart's group[23-25]has synthesized a new compound, named $ExBox^{4+}$. They have explored the ability of $ExBox^{4+}$ to selectively bind poly aromatic hydrocarbons (PAHs). Hence $ExBox^{4+}$ is capable of removing highly toxic and carcinogenic PAHs from water supplies and other mixtures. A recent theoretical report[26] has extensively studied the structure and energetics of $ExBox^{4+}$ and its host-guest complexes by means of DFT. The compound contains pyridinium-phenyl-pyridinium chains at the top and bottom. Several previous stuydies[27,11] reported gas trapping ability of aromatic ring compounds. These studies encouraged us to investigate the hydrogen adsorption capacity of this newly synthesized compound $ExBox^{4+}$since the compound contains several aromatic six member rings considering the two important problems being faced by humankind, viz., energy and environment, we have chosen adsorption of $H_2$ and CO respectively. We also study the gas adsorption ability of the Li doped $ExBox^{4+}$, since it has been already revealed that Li doped aromatic hydrocarbons can act as effective gas trapping material[11,28]as well as Li doping improves the enthalpy of adsorption and adsorption energy by a significant amount. For this purpose we place eight Li atoms above the six membered rings, phenyl and pyridinium rings, and then investigate the change in interaction energy and enthalpy as well as the gas adsorption capacity on metal doping.

**Computational Details:**

The adsorption of gas molecules is generally driven by the non-covalent van der Waals interaction or electrostatic interaction. In order to get a compact geometry and good interaction energy the optimization and frequency calculation are performed using wB97x-D method. The wB97x-D is the latest functional from Head-Gordon and coworkers,[28] which includes empirical dispersion as well as long range corrections. The structures of $ExBox^{4+}$ and its gas adsorbed analogues are optimized at the wB97x-D/6-311G (d, p) level of theory. In order to optimize the effect of Li doping on the hydrogen adsorption capacity and the adsorption energy and reaction enthalpy, the $ExBox^{4+}$ is decorated with Li and then hydrogen adsorption is studied on the Li decorated $ExBox^{4+}$ compound. The structures of $ExBox^{4+}$, $Li_8ExBox^{4+}$ and the gas adsorbed analogues are optimized at the wB97x-D/6-311G(d,p) level of theory. To observe the dependence of geometry and adsorption energy, and thermodynamics on the method of computation for adsorption process the geometry optimization as well as frequency calculation are performed at M06/6-311G(d, p). The hybrid meta-GGA (M06)[29]functional accounts for non-covalent attractions and dispersion. The NPA charges are calculated for all the systems.

Molecular dynamics (MD) simulations are performed using atom-centered density matrix propagation (ADMP).[30] The time step for ADMP-MD simulation is set at 2fs. The energies and molecular property of all the systems for simulating trajectories are calculated at the wB97x-D/6-311G (d, p) level of theory and then the conformation allowed propagating within the ADMP approach. For thermostatic simulation the temperature is maintained using velocity scaling method during the course of MD-simulation. The G09 package[31] is used for above calculations.

AIM analysis is performed for model systems. The topological analysis are carried out on the optimized geometries of the studied systems, by exporting a wfn file from GAUSSIAN09 program generated at the wB97x-D/6-311G (d, p) level and introducing into AIM2000.[32]

**Results and discussion:**

This section is divided into a few subsections, first the molecular hydrogen and CO adsorption on the $ExBox^{4+}$ molecule is discussed, followed by the effect of Li doping on hydrogen adsorption as well as CO adsorption is discussed. The adsorption energy, reaction enthalpy are calculated for the hydrogen adsorption on the $ExBox^{4+}$ molecule, the nature of bonding is studied with the help of AIM analysis. For the gas adsorption on Li doped $ExBox^{4+}$ molecule the adsorption energy, reaction enthalpy are calculated followed by ADMP simulation study. The nature of bonding is also studied with the aid of AIM2000.

**Gas adsorption ability of $ExBox^{4+}$**

***Hydrogen adsorption***

The $ExBox^{4+}$ molecule (Figure 1) contains two pyridnium-phenyle-pyridinium chains at the top and bottom. $ExBox^{4+}$ can adsorb hydrogen molecule both in exohedral and endohedral manner. Here up to three hydrogen molecules adsorption in endohedral manner is studied. For exohedral hydrogen adsorption, at first eight hydrogen molecules are placed just above the eight six membered rings resulting in $(8H_2)_{exo}$@$ExBox^{4+}$ and is then optimized. The optimized geometry belongs to minimum on the potential energy surface (NIMAG=0) and $C_1$ point group. In the next step four more hydrogen molecules are placed at the four corners of the $ExBox^{4+}$ resulting in $(12H_2)_{exo}$@$ExBox^{4+}$ compound, which is then optimized. Here also the optimized geometry belongs to a minimum on the potential energy surface. The endohedral hydrogen traping on the $(8H_2)_{exo}$@$ExBox^{4+}$ system results in a complex attached to more hydrogen, i.e. $(8H_2)_{exo}$ +$(2H_2)_{endo}$@$ExBox^{4+}$ and $(8H_2)_{exo}$ +$(3H_2)_{endo}$@$ExBox^{4+}$. In the same way the endodedral hydrogen trapping on $(12H_2)_{exo}$@$ExBox^{4+}$ further gives a complex with two more hydrogen trapped complex $(12H_2)_{exo}$ +$(2H_2)_{endo}$@$ExBox^{4+}$ and $(12H_2)_{exo}$ +$(3H_2)_{endo}$@$ExBox^{4+}$. This makes $ExBox^{4+}$ molecule to be proficient of loading up to 4.3wt% of hydrogen. Its hydrogen adsorption capacity is comparable to that of the zeolite imidazolate frameworks (ZIF-8, ZIF-11, ZIF-95, and ZIF-100).[35-35]

The average adsorption energy per molecule of hydrogen adsorbed is calculated using the following equation,

$$E_{ads} = \frac{\left(E_{nH_2@ExBox^{4+}} - (E_{ExBox^{4+}} + nE_{H_2}\right)}{n} \ldots\ldots\ldots\ldots (1)$$

where n is the number of adsorbed hydrogen molecules.

The average adsorption energy for endohedral and exohedral hydrogen adsorption (Table 1) suggests that endohedral hydrogen molecules interact more strongly with the parent moiety than the exohedral one. The hydrogen adsorption energy seems to be close to that of the recently studied charged fullerene.[16b]The variation of adsorption energy in Figure 2 interestingly suggests that the adsorption energy for simultaneous exohedral and endohedral hydrogen adsorption is higher (absolute value) than the merely exohedral hydrogen adsorption. This may be because of the endohedral hydrogen molecules which increase the overall hydrogen adsorption energy. If only exohedral hydrogen adsorption is considered, then it can be noted that with increasing the number of hydrogen molecules around the parent moiety the adsorption energy decreases (absolute value), as in the case of $(8H_2)_{exo}$@$ExBox^{4+}$ and $(12H_2)_{exo}$@$ExBox^{4+}$compounds, because of increasing steric crowding. The process of hydrogen adsorption is exothermic in nature. The reaction enthalpy value for endohedral hydrogen adsorption is higher than that the exoherdal adsorption process (Table 1). Further the exothermal hydrogen adsorption suggests that the hydrogen adsorption process expected to be thermodynamically spontaneous at somewhat low temperature at which the $\Delta H$ term dominates over the $-T\Delta S$ term.

The HOMO-LUMO gap (HLG) for the $ExBox^{4+}$ is 7.689ev and that of its hydrogen trapped analogues varies in the range of 7.698-7.746eV (Table 1). The hydrogen adsorption improves the HLG compared to that of bare $ExBox^{4+}$. The systems with endohedrally adsorbed hydrogen molecule possess higher HLG than that of the system with exohedrally bound hydrogen molecules. Endohedral hydrogen trapping is stronger than the exohedral trapping as suggested by HLG. Just as in the case of adsorption energy, the HLG also exhibits the same trend, i.e. the HLG for simultaneous exohedral and endohedral hydrogen adsorption is higher than the merely exohedral hydrogen adsorption. Figure 2 shows that the systems with high value of adsorption energy also exhibit high HLG.

To build a general idea for the nature of interaction between the hydrogen molecule and the adsorbent, AIM analysis of the hydrogen bound structures is performed. The AIM analysis of the compound with enedohedrally adsorbed hydrogen molecules suggests that there are non-covalent interactions operative between hydrogen molecules and the parent moiety. The AIM analysis of $(2H_2)_{endo}@ExBox^{4+}$suggests that adsorbed endohedral hydrogen molecules interact with the pyridinium ring (mostly with the N atom) and the phenyl ring (mostly with the C atom) of the side chain (Figure S1). The bonding interaction is mostly of van der Waals type or weak ion-dipole type interaction. But there is no interaction between the adsorbed hydrogen molecules observed. The introduction of third hydrogen molecule (resulting in $(3H_2)_{endo}@ExBox^{4+}$ compound) slightly changes the topology (Figure S1). The third hydrogen molecule is found to be involved in non-covalent interaction with the phenyl rings of the top and bottom chains as well as with the other two hydrogen molecules. The exohedral hydrogen adsorption is studied on a model system. In a recent DFT study[26] Figure S2a, model, is used to mimic the acyclic reference for the tri-aryl top and bottom of $ExBox_4^+$. First a single hydrogen molecule is placed only on each six membered ring to mimic the structure of $(8H_2)_{exo}@ExBox^{4+}$(Figure S2b). It is observed that hydrogen molecules are held by weak van der Waals force or electrostatic interaction (Figure S3). Further two more hydrogen molecules are added to mimic the structure of $(12H_2)_{exo}@ExBox^{4+}$(Figure S2c). The hydrogen molecules are found to interact with the pyridine ring and with the carbon atom of terminal ($-CH_3$) group through weak van de Waals/ weak electrostatic interactions (Figure S3). It is observed that the hydrogen molecules also interact among themselves through non-covalent interactions (Table S1).

### *CO adsorption:*

It is observed that $ExBox^{4+}$ can adsorb up to two endohedral molecules resulting ***in*** $(CO)_{endo}@ExBox^{4+}$and $(2CO)_{endo}@ExBox^{4+}$as products (Figure S4) with adsorption energies of -3.5kcal/mol and -6.0kcal/mol respectively. The CO adsorption process is exothermic in nature (ΔH=-2.5kcal/mol and -5.0kcal/mol). Any further addition of endohedral CO molecule results in a geometry with three imaginary frequencies, on twisting the third CO molecule along the direction of mode of vibration it is observed that the third CO molecule starts to move away from the cage although the CO molecules remain adsorbed with adsorption energy of -5.5kcal/mol. The CO adsorption improves the HLG compared to that of bare $ExBox^{4+}$. HLG for $(CO)_{endo}@ExBox^{4+}$, $(2CO)_{endo}@ExBox^{4+}$ and $(3CO)_{endo}@ExBox^{4+}$ complexes are 7.697eV, 7.734eV and 7.739 eV respectively. Successive CO adsorption improves the HLG of the complex. The AIM analysis of the compound with endohedrally adsorbed CO molecules suggests that there are non-covalent interactions operative among hydrogen molecules and parent moiety as portrayed by corresponding ρ, $\nabla^2 q$, and $H_{rcp}$ values (Table S2). In case of $(CO)_{endo}@ExBox^{4+}$ complex it is observed that the CO molecule interacts with the phenyl ring of the top and bottom chains. It is also observed that it interacts with the C atoms of the phenyl ring of one chain and with the hydrogen atoms of the other chain. In case of $(2CO)_{endo}@ExBox^{4+}$, the CO molecules interact with the pyridinium rings of top and bottom chains and also with the phenyl rings of the side chains. The carbon end of CO molecules is found to interact with the nitrogen atom of the pyridinium ring. For $(3CO)_{endo}@ExBox^{4}$ the endohedral CO molecules interact with each other (Figure S4) via a non-covalent interaction as suggested by corresponding ρ, $\nabla^2 q$, and $H_{rcp}$ values (Table S2).

No minimum energy structure is found with exohedral CO adsorption on $ExBox^{4+}$ at the studied level of theory.

**Effect of Li doping on gas adsorption:**

The reported[26] lowest energy $C_{2v}$ structure is considered for Li doping. In order to achieve the highest storage capacity 8 Li atoms are placed just above the pyridine rings and on the phenyl rings. The Li doped $ExBox^{4+}$ ($Li_8ExBox^{4+}$) belongs to $C_{2h}$ point group (Figure 3). Both the wB97Xd and M06 optimized geometries belong to the $C_{2h}$ symmetry. The calculated binding energy at the wB97Xd/6-311G(d,p) level is -40.7 kcal/mol, which is higher than the cohesive energy of Li in bulk phase. The large binding energy prevents the possibility of Li clustering problem. The reaction enthalpy ($\Delta H$ = -41.4kcal/mol) value suggests that the interaction of Li with $ExBox^{4+}$ is exothermic in nature. The reaction free energy value ($\Delta G$ = -33.4kcal/mol) indicates that doping process is of thermodynamically spontaneous. The calculated charge on the Li center varies in the range of 0.897-0.943 au computed with wB97Xd method (0.901-0.939 au for M06 method). The analysis of NPA charges suggests that the Li centers above the phenyl ring of the side chains carries higher charge than the other.

***Hydrogen adsorption:***

The structure of the Li doped cage remains almost unaltered on adsorption of a single hydrogen molecule by each Li center (Figure 4 a) resulting $8H_2$@ $Li_8ExBox^{4+}$ compound. Further adsorption of hydrogen molecule to give $16H_2$@ $Li_8ExBox^{4+}$also leaves the lithium doped cluster unaltered (Figure 4b). But addition of third hydrogen around each metal center slightly distorts the Li doped cage even though Li center effectively holds three hydrogen molecules (Figure 4c). There is a very little effect of successive hydrogen adsorption on H-H bond length observed in both methods of computation. The method of computation has no effect on $d_{H-H}$ value. The bond length of adsorbed hydrogen is ~0.75Å, there occurs an elongation in H-H bond due to charge redistribution. The Li-H bond lengths are listed in Table 2. It is observed that the strength of Li-H interaction is inversely proportional to their respective Li-H bond lengths. The ranges of Li-H bond lengths are listed in Table 2. The Li-H bonds computed through the M06 method are slightly elongated than that using the wB97x-D method. In both the methods the Li-H bond distance increases on successive adsorption of hydrogen molecules. The <HHLi bond angle varies in the range 80-89° in the hydrogen adsorbed complexes. In most hydrogen adsorbed complexes $H_2$ molecule prefers to stick to the Li center leaning slightly towards the ─C═C─ unit of the corresponding six membered ring. The variation of charge on Li center with successive hydrogen adsorption is listed in Table 2. The successive hydrogen adsorption causes decrease in the charge on the Li centers as expected.

In order to understand the nature of adsorption of gas molecules, the adsorption energy is calculated using the following equation for both wB97x-D and M06 methods,

$$E_{ads} = \frac{\left(E_{\mathrm{nM@\,Li_8ExBox^{4+}}} - \left(E_{\mathrm{Li_8ExBox^{4+}}} + nE_M\right)\right.}{n} \dots\dots\dots\dots (2)$$

where n is the number of adsorbed gas molecules and M= $H_2$ and CO.

Some previous studies revealed that van der Waals interaction plays an important role in hydrogen adsorption process. In order to consider van der Waals interaction, the adsorption energy is calculated using wB97x-D method which includes empirical dispersion as well as long range terms. For comparison M06 method is used for adsorption energy calculation. The adsorption energy exhibits the same trend using wB97x-D and M06 methods (Figure S5). It is observed that adsorption energy per mol $H_2$ adsorbed decreases (absolute value) with successive adsorption of hydrogen molecules. The hydrogen adsorption energy falls in the optimal range of 4.8-4.3 kcal/mol. Each Li center holds three hydrogen molecules resulting in 6.23 wt% of hydrogen storage capacity, which meets the DOE target. The adsorption energy calculated at the wB97x-D/6-311G(d, p) level is close to that obtained at the M06/6-311G(d, p) level. Both the methods, wB97x-D and M06, exhibit very close adsorption energy values for $24H_2@Li_8ExBox^{4+}$ complex (Figure S5). The reaction enthalpy values suggest the hydrogen adsorption processes is exothermic in nature (Table 2).The metal doping increases the adsorption energy and reaction enthalpy by ~3kcal/mol. There also occurs a remarkable increase in hydrogen storage capacity, the storage capacity becomes almost comparable with that using the MOF-5.[36]

The variation in reaction free energy (ΔG) with temperature is calculated with the wB97x-D and M06 methods. It is observed that ΔG value decreases with temperature. The ΔG vs T plot (Figure 5a) for adsorption of eight $H_2$ molecules resulting in $8H_2@$ $Li_8ExBox^{4+}$ complex exhibits same trend of ΔG in both methods of calculation. The M06 method underestimates the ΔG value. It is observed that below 175K the adsorption process becomes thermodynamically spontaneous (wB97x-D method). The ΔG vs. T plots for further adsorption of hydrogen around Li center ($16H_2@$ $Li_8ExBox^{4+}$ and $24H_2@$ $Li_8ExBox^{4+}$ complexes) exhibit that the ΔG values drop linearly with temperature for the wB97x-D method, whereas for M06 method a vale is observed near 150K (Figure 5b and 5c). For $16H_2$ and $24H_2$ adsorbed systems somewhat lower temperature is required to achieve thermodynamic spontaneity.

*ADMP simulation study:*

The time evolution of $8H_2@$ $Li_8ExBox^{4+}$ is examined by taking the geometry obtained using the wB97x-D method. The simulation started from the optimized geometry. In this present simulation two different conditions are applied, adiabatic and thermostatic MD. Thermostatic simulations are performed at 298K, 100K, and 77K temperatures.Time evolution of relative energy for adiabatic and thermostatic MD simulations during 600fs (except for 298K) is shown in Figure 6a. Energy is more or less conserved during the adiabatic process. The energy profiles at temperatures 100K and 77K are similar but that of 298 K is different, the amplitude of oscillation is higher (shown in inset). It is observed that for 100K, and 77K temperatures throughout the simulation time all eight $H_2$ molecules remain bound to the host molecule but for 298K simulation one $H_2$ molecule starts moving away from the Li center (Li96-104H) within 40fs. It can be seen from Figure 7 that at 298K, within 150fs three $H_2$ molecules are desorbed from the parent moiety. On the contrary, for 100K and 77K simulations all eight hydrogen molecules remain adsorbed on parent moiety, they only change their orientation around the metal center. The parent cage gets slightly distorted in thermostatic simulation as well as adiabatic simulation. The time evolution of 98Li-102H and 96Li-102H bond length for adiabatic and 100K simulation exhibit that with time the hydrogen molecule attached to 98Li moves closer to 96Li, after 600fs it gets adsorbed on 96Li. No such phenomenon is observed for simulation at 77K, the time evolution of shortest Li-H bond length (oscillates in the range 2-2.5Å) is shown in Figure S6. The time evolution of H-H bond length is shown in Figure S6. It is observed that hydrogen remains adsorbed in molecular form and H-H bond length oscillates around 0.75Å. The time evolution of H-H bond length is very similar for both adiabatic and thermostatic (298K,100K, and 77K) MD simulation. The amplitude of oscillation increases with temperature. The HLG profile shows that throughout the simulation time the hydrogen bound complex maintains a substantial HLG which is required for the stability of the system.

The time evolution of $16H_2$@ $Li_8ExBox^{4+}$ is examined by using wB97x-D method at three different temperatures, 298K, 100K and 77K by means of thermostatic ADMP-MD simulation. In order to reduce computational cost thermostatic MD simulations are performed for 300fs for 100K and 77K. For 298K, the simulation is performed for 100fs, since within this time there occurs a considerable change in the structure of the studied compound. Figure 6b exhibits time evolution of relative energy for all thermostatic MD simulations during 300fs (except for 298K). The energy profile for temperature 100K and 77K are extremely similar but that of 298 K is different, the amplitude of oscillation is higher. It is observed that at 298K, within 50fs only one hydrogen molecule desorbed from the parent moiety. At 100fs three hydrogen molecules get desorbed from the parent moiety and other hydrogen molecules reorient around the metal centers (Figure 8). On the other hand it is observed that up to 300fs all sixteen $H_2$ molecules retained in thermostatic simulation at both 100K and 77K temperatures(Figure 8) and the shortest Li-H bond length oscillates in the range 2-2.5Å (Figure S7). A little deformation of the parent cage in thermostatic simulation is observed. The time evolution of H-H bond length is shown in Figure S7. The hydrogen remains adsorbed in molecular form with an H-H bond length around 0.75Å. The HLG profile shows that throughout the simulation time the hydrogen bound complex maintains a substantial HLG required for.

For $24H_2$@ $Li_8ExBox^{4+}$, the thermostatic ADMP-MD simulations are performed at 298K and 77K temperatures. As in the case of 16 $H_2$ adsorbed system the simulation at 77K is performed for 300fs. Due to considerable change in structure the simulation at 298K is performed for only 100fs. Figure 6c exhibits time evolution of relative energy for thermostatic MD simulations at 298K (during 100fs) and 77K (during 300fs). It is observed that at 298K, within 60fs one hydrogen molecule per Li center starts to move away and within 100fs 10 hydrogen molecules get desorbed from the parent moiety (Figure 8). The thermostatic MD simulation of $24H_2$@ $Li_8ExBox^{4+}$ complex reveals that all 24 $H_2$ molecules retained up to 300fs at 77K temperature. The hydrogen remains adsorbed in molecular form and H-H bond length oscillates around 0.75Å (Figure S8). The HLG profile shows that throughout the simulation time the hydrogen bound complex maintains a substantial HLG which is required for the stability of the system (Figure S8).

The ADMP simulation study reveals that Li doped $ExBox^{4+}$ is capable of holding 24 hydrogen molecules at a low temperature for considerably long time. At room temperature the hydrogen molecules starts to desorb from the parent moiety with in 100fs. Hydrogen remains adsorbed in molecular form and the hydrogen adsorbed complexes remains stable throughout the simulation as suggested by the HLG of the respective systems even at room temperature. The simulation study suggests that the complex is capable of hydrogen storage at optimum temperature, as is also suggested by their free energy plot.

***CO adsorption:***

In this section the CO binding ability of lithium decorated $ExBox^{4+}$ is discussed. In the first step eight CO molecules are introduced to the $Li_8ExBox^{4+}$ cluster. The optimized geometry of 8CO@ $Li_8ExBox^{4+}$ cluster is given in Figure9a. Just like hydrogen adsorption process both

wB97x-D and M06 methods in combination with 6-311G(d, p) basis set are used for geometry optimization and frequency calculation. All the reported structures belong to minima on the potential energy surface. The Li-C bond lengths are listed in Table 3. It is noted that there is a good agreement between the two methods of computation in terms of Li-C bond length. The <OCLi bond angle varies in the range of 177º-180º. In 8CO@ $Li_8ExBox^{4+}$ cluster C-O bond length is 1.11754Å, slightly shorter than in a free CO molecule. It is interesting to note that on addition of eight more CO molecules each Li center binds two CO molecules, but it is observed that the second CO molecule on the Li center above pyridine ring binds the CO molecule in such a manner that the oxygen end of the CO gets bonded with the vicinal Li center (above the phenyl ring) resulting in an increase in total bonding interaction (Figure 9b). This unique bonding structure results in a shorter Li-C bond, even shorter than that in 8CO@ $Li_8ExBox^{4+}$. The CO bond length, for the CO bonded with C end is 1.11762 Å. The C-O bond gets elongated to 1.20196 Å, in the CO molecules in which both C and O ends are bonded to adjacent Li centers. The structure of the ExBox cage changes on adsorption of sixteen CO molecules. On the contrary to this the cage remains practically unaltered on adsorption of eight CO molecules, almost same with that of $8H_2$@ $Li_8ExBox^{4+}$ compound. The variation of charge on Li center with successive CO adsorption is listed in Table3. The successive CO adsorption results in continuous decrease in the charge on the Li centers.

The adsorption energy exhibits the same trend for wB97x-D and M06 methods (Table 3). The wB97x-D adsorption energy is quite close to that obtained from the M06 method. It is observed that each Li center holds the CO molecule quite strongly. It is interesting to note that the adsorption energy for 16CO@ $Li_8ExBox^{4+}$ is higher than that of 8CO@ $Li_8ExBox^{4+}$, this may be accounted for by unique binding patterns of two CO molecules to the Li centers. The CO adsorption process is exothermic in nature (Table 3).

The variation of reaction free energy ($\Delta G$) with temperature is calculated with the wB97x-D and M06 methods (Figure 10). It is observed that adsorption of CO molecule is thermodynamically spontaneous at room temperature. At very high temperature, ~500K the reaction free energy value is positive for adsorption of 8CO as well as 16CO molecules. For adsorption of eight CO molecules, it is observed that the reaction free energy value becomes favourable after 450K, whereas for sixteen CO molecules it becomes favourable only around ~350K. The calculated $\Delta G$ values for wB97x-D are fairly close to M06 method. The trend in $\Delta G$ values are fairly similar for adsorption of both 8CO and 16CO molecules.

*ADMP simulation study:*

Starting from the wB97x-D optimized geometry time evolution of 8CO@ $Li_8ExBox^{4+}$ is scanned in two thermostatic ADMP-MD simulations at 298K and at very low temperature, 77K for 600fs. The relative energy trajectory is given in Figure 11a. It is observed that for 298K simulation the distortion of the cage starts around 160fs, and till the end of the simulation it gets completely distorted (Figure 12). But it is worth mentioning here that till the end of the simulation all the CO molecules retained. The longest Li–C bond length oscillates between ~2.1Å to ~2.9Å (Figure S9). ADMP-MD simulation at a low temperature, 77K, explores that the distortion of cage is very little up to 600fs (Figure 12). All the eight CO molecules are retained till 600fs and the longest Li-C bond oscillates around 2.3Å. The amplitude of oscillation of C–O bond length is higher at of 298K than that at 77K (Figure S9). At 298K the C–O bond gets slightly elongated with time. For 77K simulation the C–O bond length oscillates in the range 1.1125Å to 1.1225Å. During the course of simulation the CO molecules reorient around the Li center. The amplitude of oscillation in HOMO-LUMO gap (HLG) is also higher in case of 298K simulation than that of 77K (Figure S9). The HLG varies in the range of ~0.155au to ~0.190au for 298K simulation. For 77K simulation the HLG fluctuates in the range ~0.1675au to ~0.1825au.

The time evolution of 16CO@ $Li_8ExBox^{4+}$ complex is examined through two thermostatic ADMP-MD simulations at 298K and 77K up to 300fs, starting the calculations with the wB97x-D optimized geometry. The relative energy trajectories for 298K and 77K temperatures are quite similar (Figure11b). It is observed that for 298K simulation the distortion of the cage is detected, but the extent of cage distortion is lower than that in the case of 8CO@ $Li_8ExBox^{4+}$. For 298K simulation it is observed that at least one CO molecule from the Li center above the middle phenyl rings (of the top and bottom chain) starts to move away after 200fs. Till the end of the simulation fourteen CO molecules remain within the bonding distance of the parent moiety (Figure12). ADMP-MD simulation at a low temperature, 77K, explores that the distortion of cage is very little up to 300fs. All the sixteen CO molecules retain till 300fs (Figure12). The C─O bond length oscillates between ~1.10Å-1.14Å and ~1.115Å-1.13Å for simulation at the temperatures 298K and 77K respectively (Figure S10). The HLG decreases with time for 298K, whereas the HLG oscillates around ~0.20 au for 77K simulation (Figure S10).

The ADMP simulation reveals that at room temperature the complex can adsorb up to fourteen CO molecules at room temperature till the end of simulation time, whereas at lower temperature all sixteen CO molecules remains adsorbed through out the simulation. The 16CO@ $Li_8ExBox^{4+}$ complex at becomes unstable with time at 298K, as suggested by the HLG profile. The system remains stable during the simulation at 77K. The ADMP simulation suggests that the complex can be used as CO storage material at room temperature as well as low temperature.

**Nature of Interaction:**

In a recent DFT study[26] a model (Figure S2a) is used to mimic the acyclic reference for the tri-aryl top and bottom of $ExBox_4^+$. Here we have considered Li doped model with 2 unit of positive charge as model system and then added three hydrogen and carbon monoxide molecules (Figure 13). In order to inspect the nature of bonding between the Li center and $H_2$ and CO molecules AIM analysis is performed on gas adsorbed model system. The model system and the gas adsorbed analogues are first optimized at the wB97x-D/6-311G(d, p) level. The topological analysis is carried out on the optimized geometries using AIM2000 at same level by exporting a wfn file from GAUSSIAN09 generated at the wB97x-D/6-311G(d, p) level and introducing into AIM2000. FigureS11 displays the molecular graphs of $3H_2$@$Li_3$model and 3CO@$Li_3$model. The $\rho_{BCP}$ and $\nabla^2\rho_{BCP}$ values at the BCP of the Li–H bond are 0.00964 au and 0.05268au respectively, clearly indicating the existence of ionic or van der Waals interaction between these two atoms (Table 4). The positive value of local energy density (0.00314 au) also indicates the same. It is also observed that for the $H_2$ above the pyridine rings there is a weak bonding interaction between the carbon atom and adsorbed hydrogen molecule as there exists a BCP between C and H atoms. The $\rho_{BCP}$(0.00576 au) and $\nabla^2\rho_{BCP}$(0.01179 au) values suggest that the bond is of weak van der Waals type. The topological analysis of 3CO@ $Li_3$model suggests (Table 4) that the Li–C interaction is also of ionic or of van der Waals. The $\rho_{BCP}$ and$\nabla^2\rho_{BCP}$ and also the local energy density value suggest that the Li-C interaction is somewhat stronger than the Li-H interaction.

## Conclusion:

The newly synthesized compound $ExBox^{4+}$ appears to be an efficient hydrogen storage material. It can hold hydrogen molecules in both exoherdal and endohedral manner. The endohedral hydrogen molecules interact strongly than exohedral ones. The hydrogen adsorption energy is comparable with that of the recently studied charged fullerenes. The hydrogen storage capacity (~4.3 wt%) appears to be as good as the zeolite imidazolate frameworks. The non-covalent interaction is found to be operative between the CO molecule and the $ExBox^{4+}$. The DFT based calculation reveals that Li doping on $ExBox^{4+}$ leads to a thermodynamically stable compound $Li_8ExBox^{4+}$. All the Li centers are positively charged. Further study on gas adsorption capacity of this Li doped compound revealed that it can efficiently bind $H_2$ and CO molecules. The careful scrutiny of the adsorption energy, reaction enthalpy and reaction free energy values suggest that under optimal temperature the adsorption of $H_2$ and CO are thermodynamically spontaneous and the resulting gas bound structures are quite stable. $Li_8ExBox^{4+}$ can act as a good $H_2$ storage material with their hydrogen storage capacity of 6.23wt%. The lithium supported on $ExBox^{4+}$is found to improve the gas storage capacity and the enthalpy value gets improved by ~3kcal/mol. The MD simulation using ADMP approach reveals that at a low temperature $Li_8ExBox^{4+}$ can hold up to $24H_2$ molecules for a fairly long period with a substantial HLG required for stability of a system. The calculated $\Delta G$ value reveals that CO adsorption process is thermodynamically spontaneous at room temperature. The MD simulations on CO adsorbed systems are performed at room temperature, 298K, and at a very low temperature, 77K, to observe the change in the dynamics with temperature. The MD simulation explores that at 298K $Li_8ExBox^{4+}$ can hold fourteen CO molecules up to 300fs with a little distortion of the cage. But at low temperature, 77K, it can hold sixteen CO molecules up to 300fs. The unique binding mode of the CO molecule, bound with both carbon and oxygen ends with the vicinal Li centers (above the phenyl ring), results in an increase in total bonding interaction. The AIM analysis on the model systems showed that the Li–H and Li–C interactions are of ionic or van der Waals type. The AIM analysis also suggests that the Li-C interaction is a bit stronger than the Li-H interaction.

## Acknowledgements:

PKC thanks the DST, New Delhi for the Sir J. C. Bose National Fellowship. RD thanks UGC, New Delhi for a fellowship.

**Table 1:** Adsorption energy ($E_{ads}$,kcal/mol), reaction enthalpy (ΔH,kcal/mol), HOMO-LUMO gap (HLG,eV) $nH_2$@ExBox$^{4+}$ calculated at wB97x-D/6-311G(d,p) basis set

| System | $E_{ads}$(kcal/mol) | ΔH(kcal/mol) | HLG(eV) |
|---|---|---|---|
| | | | |
| **$(2H_2)_{endo}$@ExBox$^{4+}$** | -2.4 | -1.047 | 7.705 |
| **$(3H_2)_{endo}$@ExBox$^{4+}$** | -2.2 | -0.896 | 7.713 |
| **$(8H_2)_{exo}$@ExBox$^{4+}$** | -1.5 | -0.217 | 7.707 |
| **$(8H_2)_{exo}$ +$(2H_2)_{endo}$@ExBox$^{4+}$** | -1.8 | -0.451 | 7.732 |
| **$(8H_2)_{exo}$ +$(3H_2)_{endo}$@ExBox$^{4+}$** | -1.8 | -0.467 | 7.739 |
| **$(12H_2)_{exo}$@ExBox$^{4+}$** | -1.5 | -0.208 | 7.698 |
| **$(12H_2)_{exo}$ +$(2H_2)_{endo}$@ExBox$^{4+}$** | -1.6 | -0.348 | 7.733 |
| **$(12H_2)_{exo}$ +$(3H_2)_{endo}$@ExBox$^{4+}$** | -1.7 | -0.359 | 7.746 |

**Table 2**: Adsorption energy ($E_{ads}$,kcal/mol), reaction enthalpy(ΔH,kcal/mol), Li─H bond length, and charge on Li center for $nH_2$@$Li_8$ExBox$^{4+}$; (n=8, 16, 24) calculated at 6-311G(d,p) basis set

| System | $E_{ads}$(kcal/mol) | | ΔH(kcal/mol) | | Li-H bond length(Å) | | $Q_{Li}$(au) | |
|---|---|---|---|---|---|---|---|---|
| | wB97XD | M06 | wB97XD | M06 | wB97XD | M06 | wB97XD | M06 |
| **$8H_2$@$Li_8$ExBox$^{4+}$** | -4.8 | -4.3 | -3.0 | -1.5 | 2.099-2.183 | 2.126-2.193 | 0.817-0.868 | 0.822-0.864 |
| **$16H_2$@$Li_8$ExBox$^{4+}$** | -4.7 | -4.3 | -3.1 | -1.9 | 2.114-2.243 | 2.113-2.274 | 0.699-0.766 | 0.702-0.768 |
| **$24H_2$@$Li_8$ExBox$^{4+}$** | -4.3 | -4.2 | -2.7 | -1.5 | 2.136-2.292 | 2.137-2.314 | 0.579-0.640 | 0.596-0.653 |

**Table 3**: Adsorption energy, reaction enthalpy, Li─C bond length, and charge on Li center for nCO@$Li_8$ExBox$^{4+}$; (n=8, 16) calculated at 6-311G (d,p) basis set

| System | $E_{ads}$ (kcal/mol) | | ΔH(kcal/mol) | | Li-C bond length(Å) | | $Q_{Li}$(au) | |
|---|---|---|---|---|---|---|---|---|
| | wB97XD | M06 | wB97XD | M06 | wB97XD | M06 | wB97XD | M06 |
| **8CO@$Li_8$ExBox$^{4+}$** | -11.4 | -11.4 | -10.3 | -10.3 | 2.274-2.307 | 2.274 -2.288 | 0.681-0.744 | 0.678-0.742 |
| **16CO@$Li_8$ExBox$^{4+}$** | -11.9 | 12.2 | -10.8 | -10.6 | 2.098-2.468 | 2.125-2.407 | 0.367-0.612 | 0.372-0.585 |
| | | | | | | | | |

**Table 4**: Bond critical point data (a.u.) calculated at wB97x-D/6-311G (d, p) level for 3M@$Li_3$model complexes; (M=CO, $H_2$)

| System | bond type | $q$ | $\nabla^2 q$ | $G_{rcp}$ | $V_{rcp}$ | $H_{rcp}$ |
|---|---|---|---|---|---|---|
| **3CO@$Li_3$model** | Li–C | 0.01578 | 0.08353 | 0.01696 | -0.01249 | 0.00447 |
| | | | | | | |
| **$3H_2$@$Li_3$model** | Li–H | 0.00964 | 0.05268 | 0.01003 | -0.00690 | 0.00314 |
| | C–H | 0.00576 | 0.01179 | 0.00250 | -0.00204 | 0.00045 |

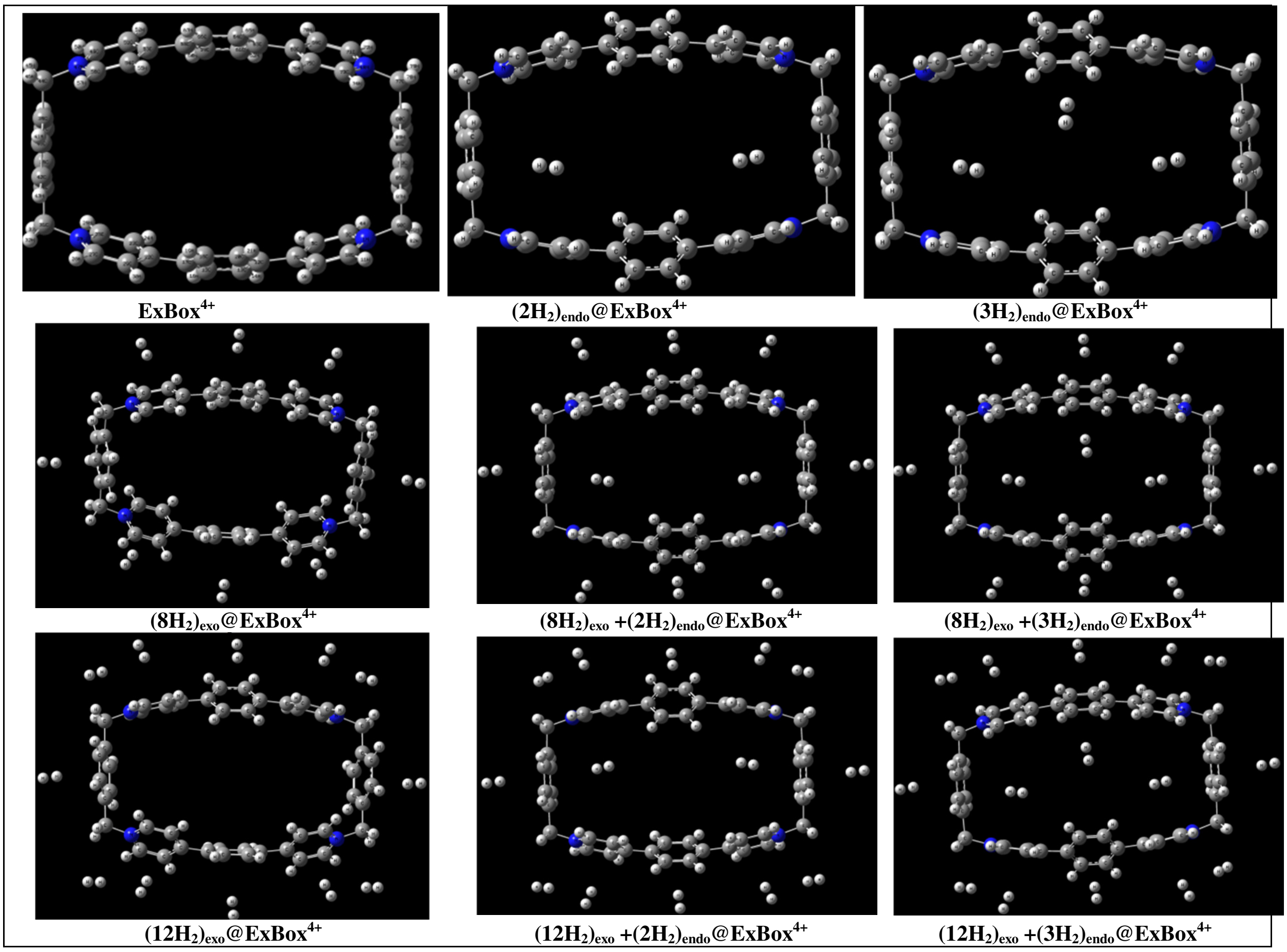

**Figure 1**: Optimized geometry of hydrogen bound $ExBox^{4+}$ complex

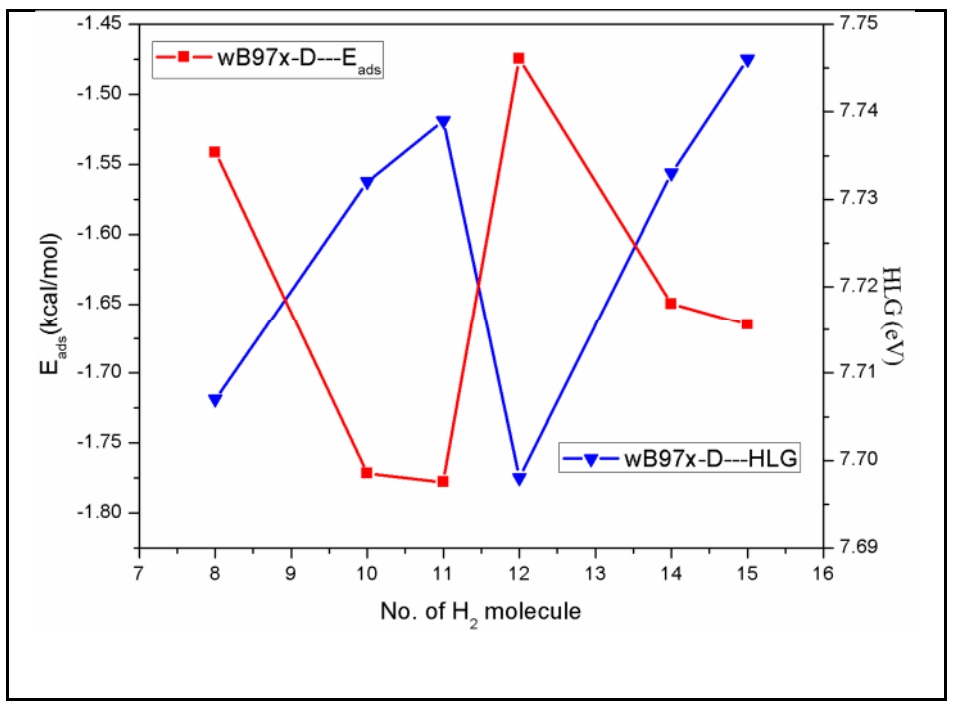

**Figure 2**: Plot of adsorption energy (Eads, kcal/mol) and HLG (eV) as a function of number of hydrogen adsorbed calculated at wB97x-D/6-311G(d,p) level.

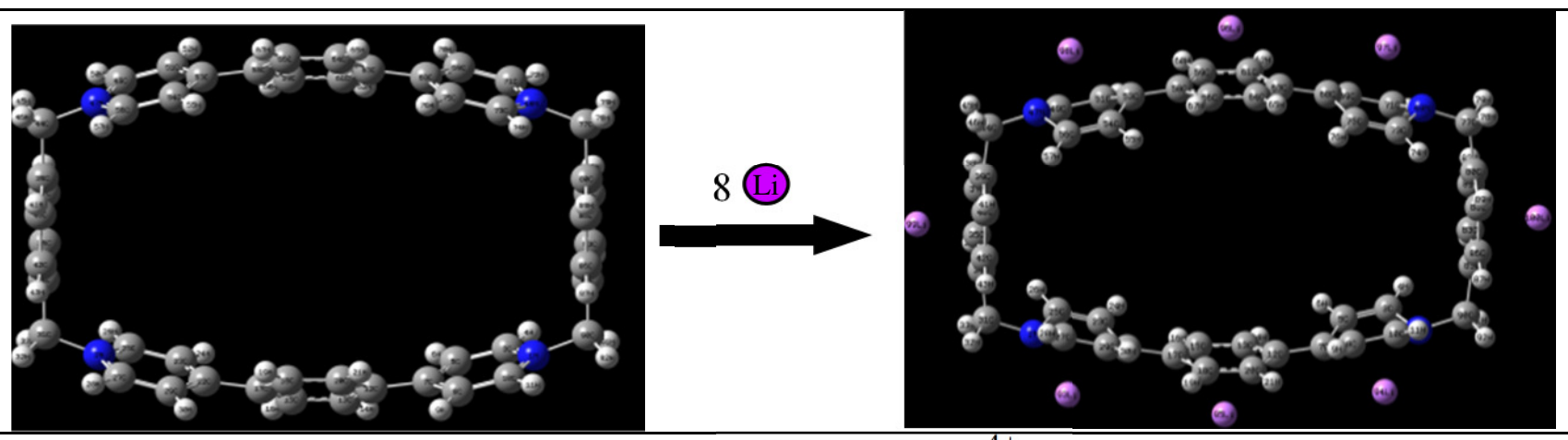

Structure of $Li_8ExBox^{4+}$

**Figure 3**: Optimized geometry of $Li_8ExBox^{4+}$ complex.

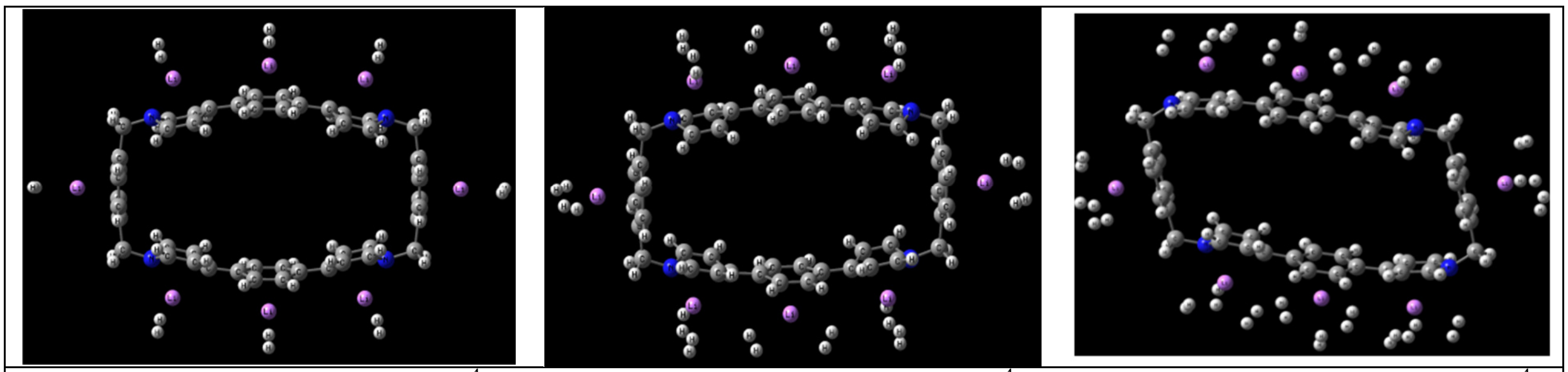

(a)Structure of $8H_2$@ $Li_8ExBox^{4+}$ (b)Structure of $16H_2$@ $Li_8ExBox^{4+}$ (c)Structure of $24H_2$@ $Li_8ExBox^{4+}$

**Figure 4**: Optimized geometry of $nH_2$@$Li_8ExBox^{4+}$ complex; n=8, 16, 24

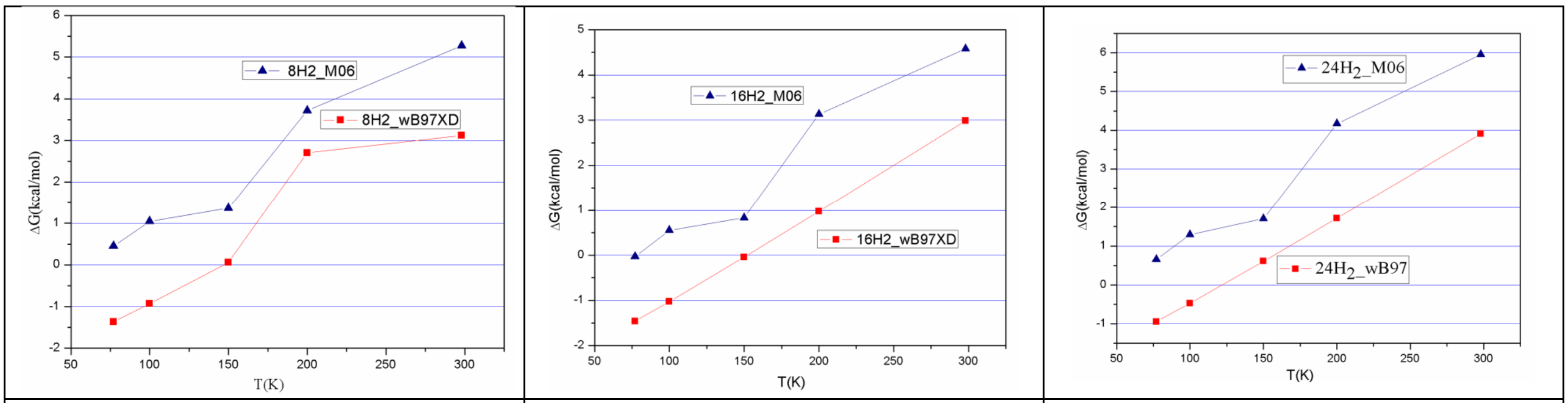

(a)$8H_2$ (b)$16H_2$ (c)$24H_2$

**Figure 5**: Variation of reaction free energy (ΔG) with temperature for the hydrogen adsorption process.

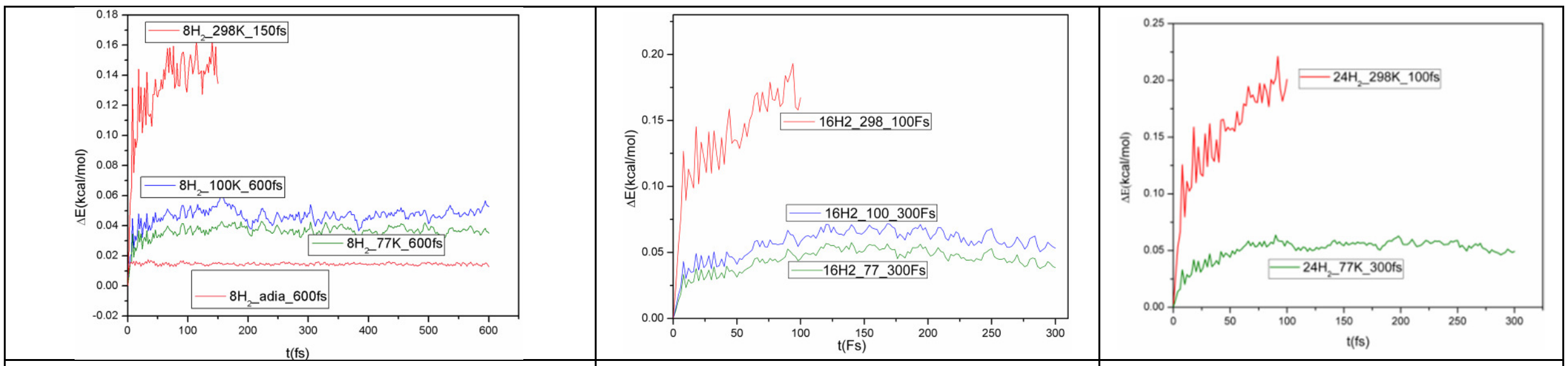

| (a) $8H_2@$ $Li_8ExBox^{4+}$ | (b) $16H_2@$ $Li_8ExBox^{4+}$ | (c) $24H_2@$ $Li_8ExBox^{4+}$ |
|---|---|---|

**Figure 6**: Relative energy trajectory for adiabatic and thermostatic simulation T= 298 K, 100 K, and 77K for $nH_2@Li_8ExBox^{4+}$ complex; n=8, 16, 24

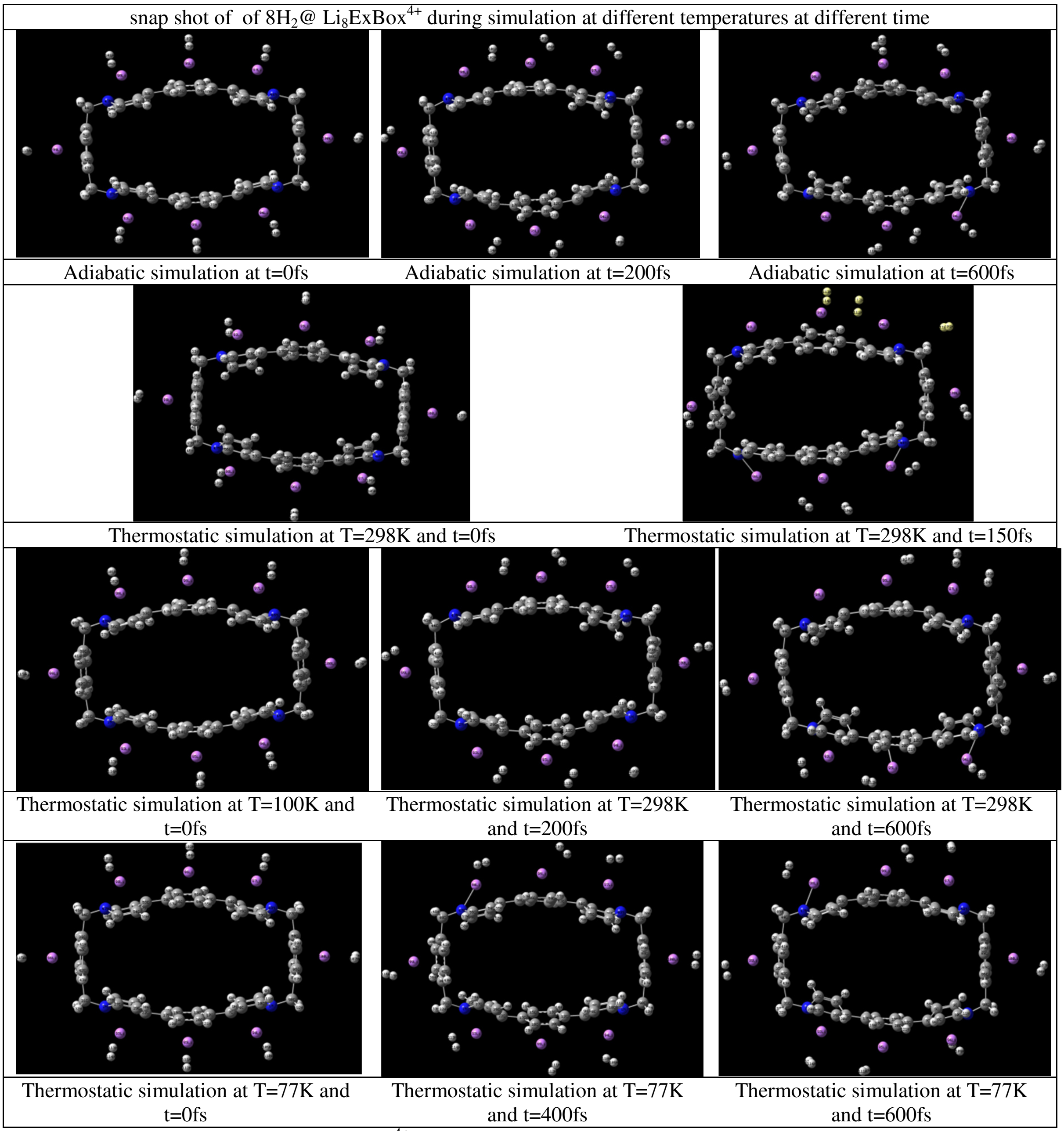

**Figure 7**: Snap shot of of $8H_2@$ $Li_8ExBox^{4+}$ under different simulation conditions at different time steps

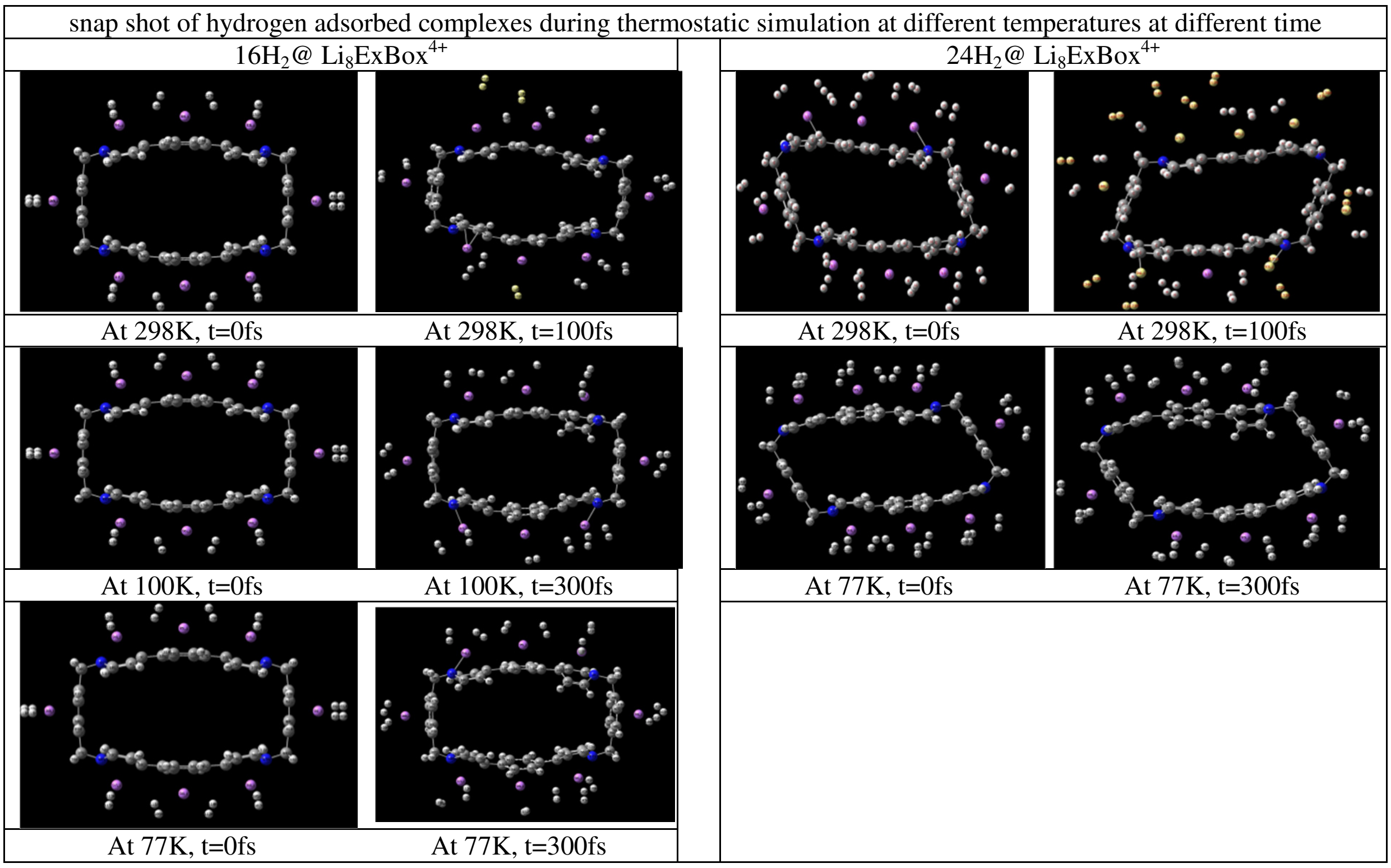

**Figure 8**: Snap shot of of $nH_2$@ $Li_8ExBox^{4+}$ complexes (n=16, 24) under different simulation conditions at different time steps

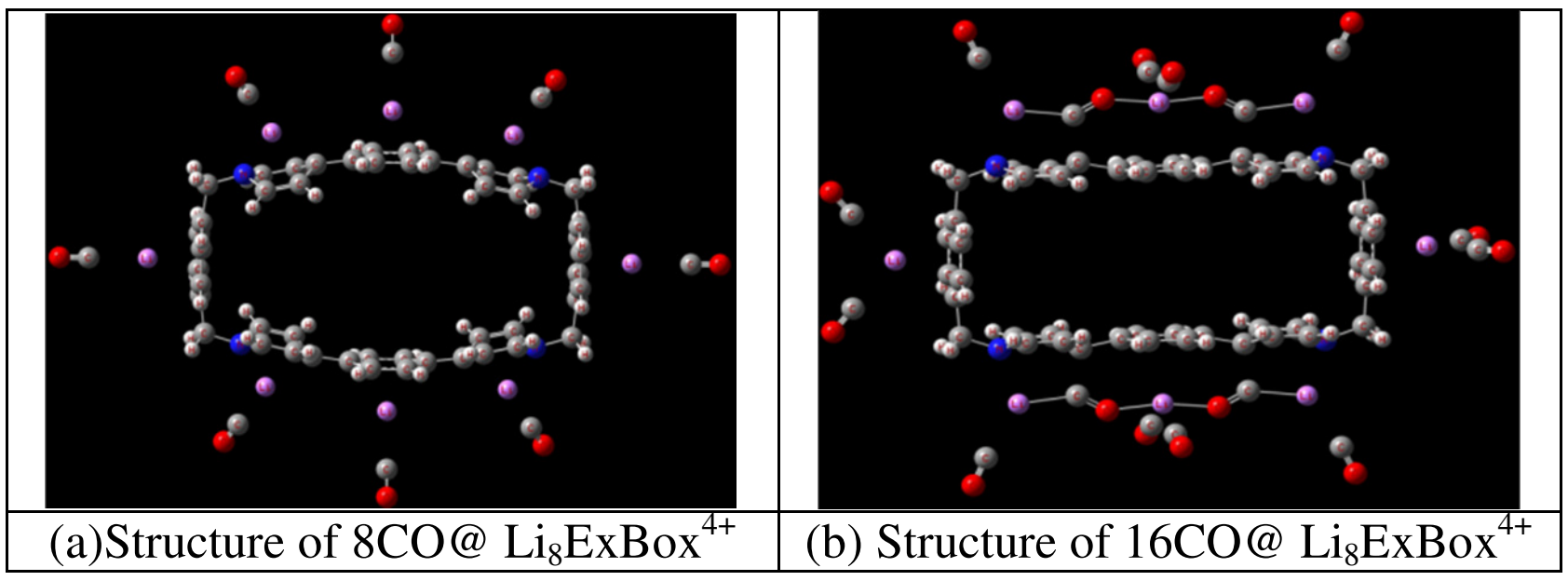

**Figure 9**: Optimized geometries of nCO@$Li_8ExBox^{4+}$ complexes; n=8, 16

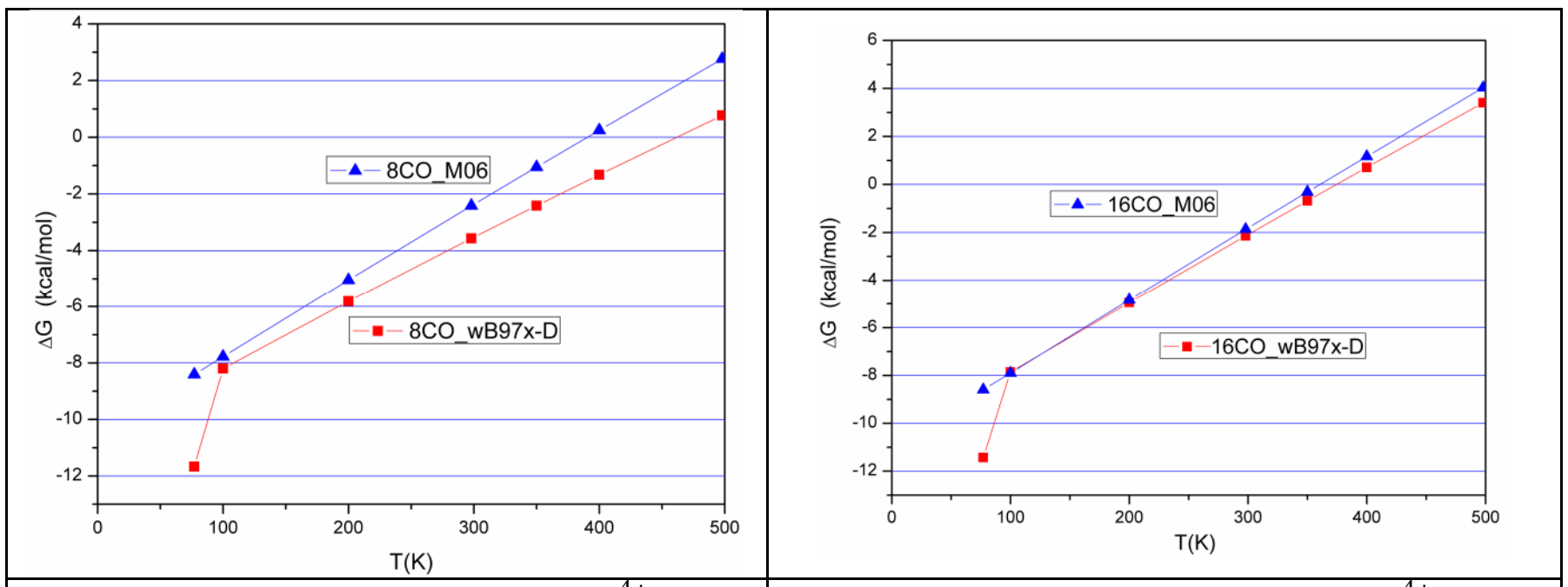

(a) 8CO@ $Li_8ExBox^{4+}$ (b) 16CO@$Li_8ExBox^{4+}$

**Figure10**: Variation of reaction free energy (ΔG) with temperature for the CO adsorption process.

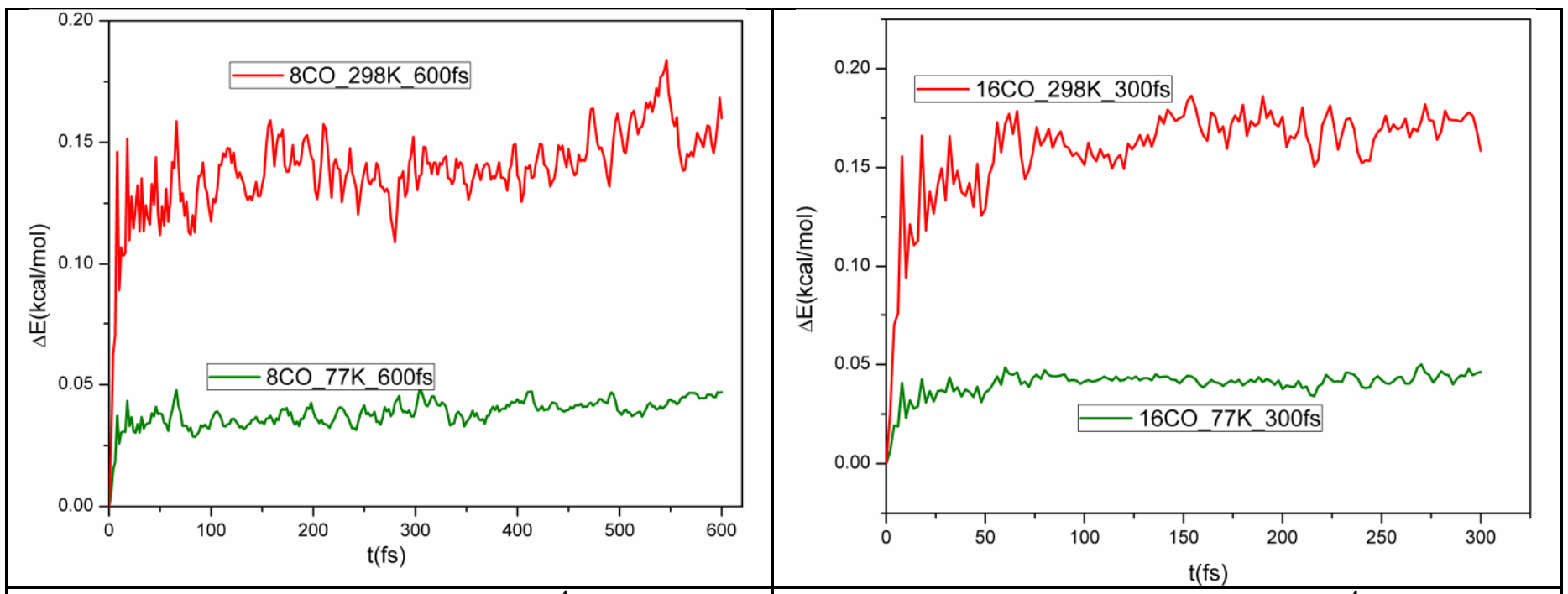

(a)8CO@ $Li_8ExBox^{4+}$ (b)16CO@ $Li_8ExBox^{4+}$

**Figure 11**: Relative energy trajectory for adiabatic and thermostatic simulation T= 298 K, 100 K, and77K for nCO@$ExBox^{4+}$ complex; n=8, 16

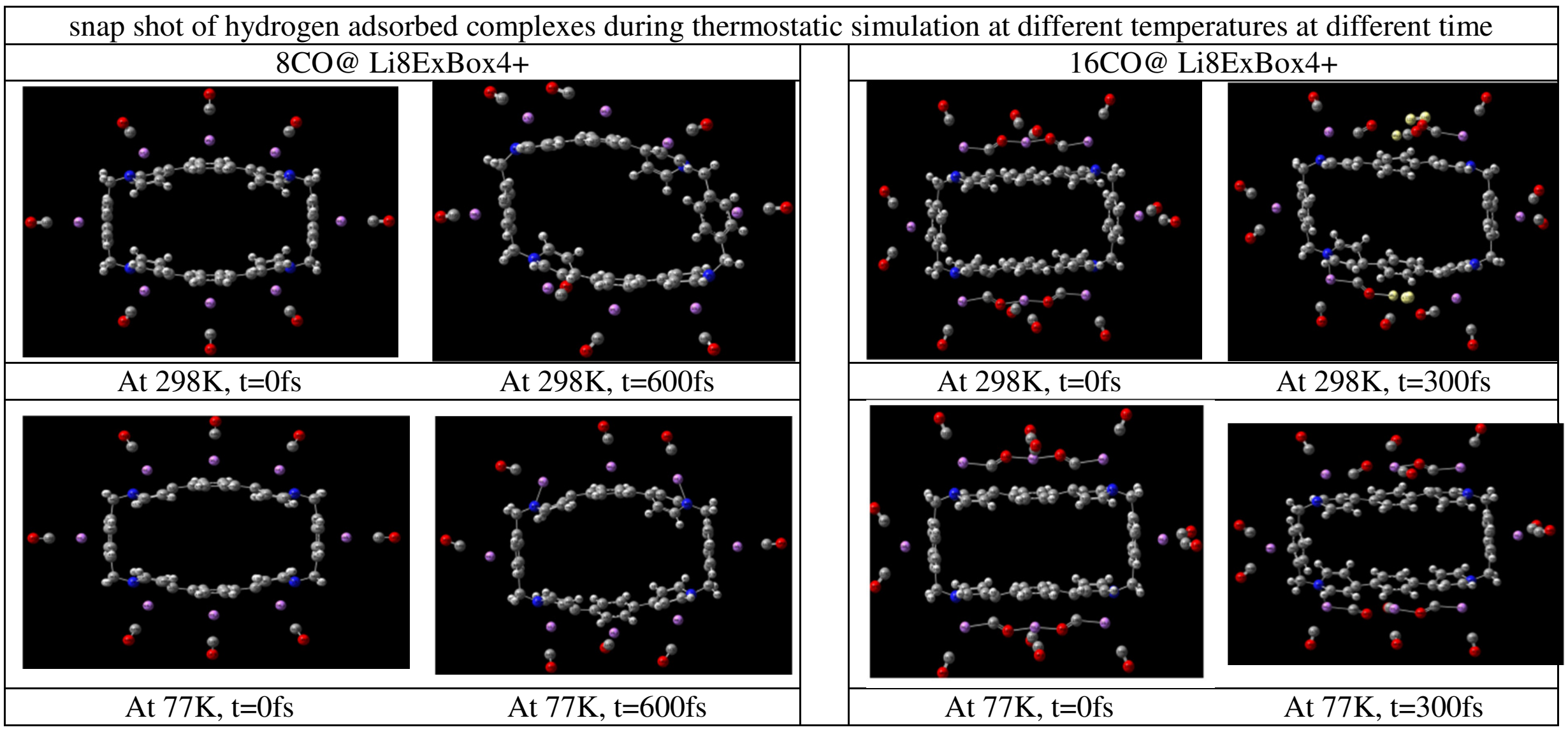

**Figure 12**: Snap shot of of $nH_2$@ $Li_8ExBox^{4+}$ complexes (n=16, 24) under different simulation conditions at different time steps

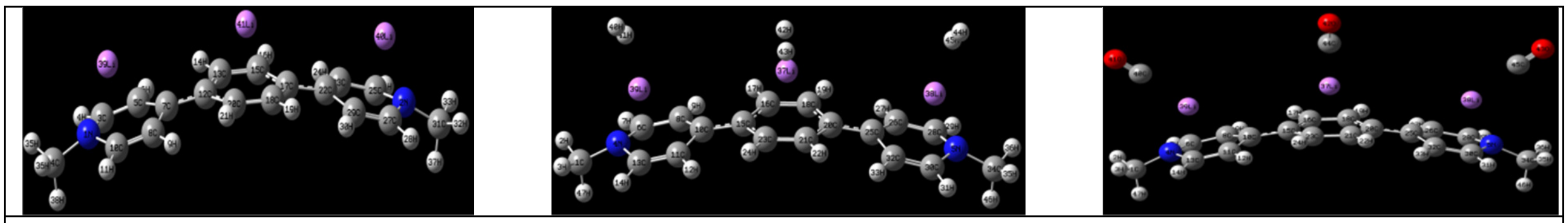

| (a)$Li_3$model | (b)$3H_2$@ $Li_3$model | (c)3CO@ $Li_3$model |
|---|---|---|

**Figure 13**: Optimized geometry at wB97x-D/6-311G (d, p) level

Supporting Information

# Gas Storage Potential of Li-decorated $ExBox_4^+$

***Ranjita Das and Pratim Kumar Chattaraj****

Department of Chemistry and Center for Theoretical Studies, Indian Institute of Technology

Kharagpur, Kharagpur – 721302, West Bengal, India

Correspondence to: Pratim K. Chattaraj (Email: pkc@chem.iitkgp.ernet.in)

Table S1:Bond critical point data (a.u.) calculated at wB97x-D/6-311G (d, p) level for hydrogen bound $ExBox^{4+}$ complexes

| System | bond type | $q$ | $\nabla^2 q$ | $G_{rcp}$ | $V_{rcp}$ | $H_{rcp}$ |
|---|---|---|---|---|---|---|
| $(2H_2)_{endo}$@$ExBox^{4+}$ | N-H | 0.00375 | 0.01304 | 0.00243 | -0.00161 | 0.00083 |
| | C-H | 0.00608 | 0.01661 | 0.00335 | -0.00255 | 0.00080 |
| | | | | | | |
| $(3H_2)_{endo}$@$ExBox^{4+}$ | N-H | 0.00385 | 0.01348 | 0.00252 | -0.00167 | 0.00085 |
| | C-H | 0.00586 | 0.01620 | 0.00325 | -0.00244 | 0.00080 |
| | H-H | 0.00025 | 0.00095 | 0.00016 | -0.00008 | 0.00008 |
| | | | | | | |
| $8H_2$@model | N-H | 0.00468 | 0.01601 | 0.00305 | -0.00209 | 0.00096 |
| | C-H | 0.00468 | 0.01526 | 0.00292 | -0.00202 | 0.00090 |
| | | | | | | |
| $12H_2$@model | C-H | 0.004055 | 0.01473 | 0.00275 | -0.00182 | 0.00093 |
| | H-H | 0.003183 | 0.01031 | 0.00192 | -0.00125 | 0.00066 |

Table S2: Bond critical point data (a.u.) calculated at wB97x-D/6-311G (d, p) level for CO bound $ExBox^{4+}$ complexes

| System | bond type | $q$ | $\nabla^2 q$ | $G_{rcp}$ | $V_{rcp}$ | $H_{rcp}$ |
|---|---|---|---|---|---|---|
| $(CO)_{endo}@ExBox^{4+}$ | C-O | 0.00402 | 0.01293 | 0.00245 | -0.00166 | 0.00079 |
| | C-C | 0.00557 | 0.01595 | 0.00316 | -0.00233 | 0.00083 |
| | H-O | 0.00068 | 0.00334 | 0.00057 | -0.00031 | 0.00026 |
| | H-C | 0.00062 | 0.00240 | 0.00041 | -0.00023 | 0.00019 |
| | | | | | | |
| $(2CO)_{endo}@ExBox^{4+}$ | C-O | 0.00417 | 0.01322 | 0.00251 | -0.00172 | 0.00079 |
| | C-O(pyridinium ring) | 0.00425 | 0.01535 | 0.00288 | -0.00192 | 0.00096 |
| | N-C | 0.00485 | 0.01597 | 0.00306 | -0.00213 | 0.00093 |
| | C-C | 0.00566 | 0.01591 | 0.00317 | -0.00236 | 0.00081 |
| | C-C(pyridinium ring) | 0.00489 | 0.01520 | 0.00294 | -0.00208 | 0.00086 |
| | | | | | | |
| $(3CO)_{endo}@ExBox^{4+}$ | C-O | 0.00409 | 0.01306 | 0.00248 | -0.00169 | 0.00079 |
| | C-C | 0.01646 | 0.14705 | 0.02757 | -0.01837 | 0.00920 |
| | third CO molecule | | | | | |
| | C-O | 0.00398 | 0.01292 | 0.00244 | -0.00165 | 0.00079 |
| | C-C | 0.00539 | 0.01545 | 0.00305 | -0.00224 | 0.00081 |
| | H-C | 0.00041 | 0.00164 | 0.00028 | -0.00015 | 0.00013 |
| | Interaction between 2CO-3CO | | | | | |
| | C-C | 0.00377 | 0.01241 | 0.00233 | -0.00156 | 0.00077 |
| | C-O | 0.00185 | 0.00799 | 0.00141 | -0.00083 | 0.00059 |

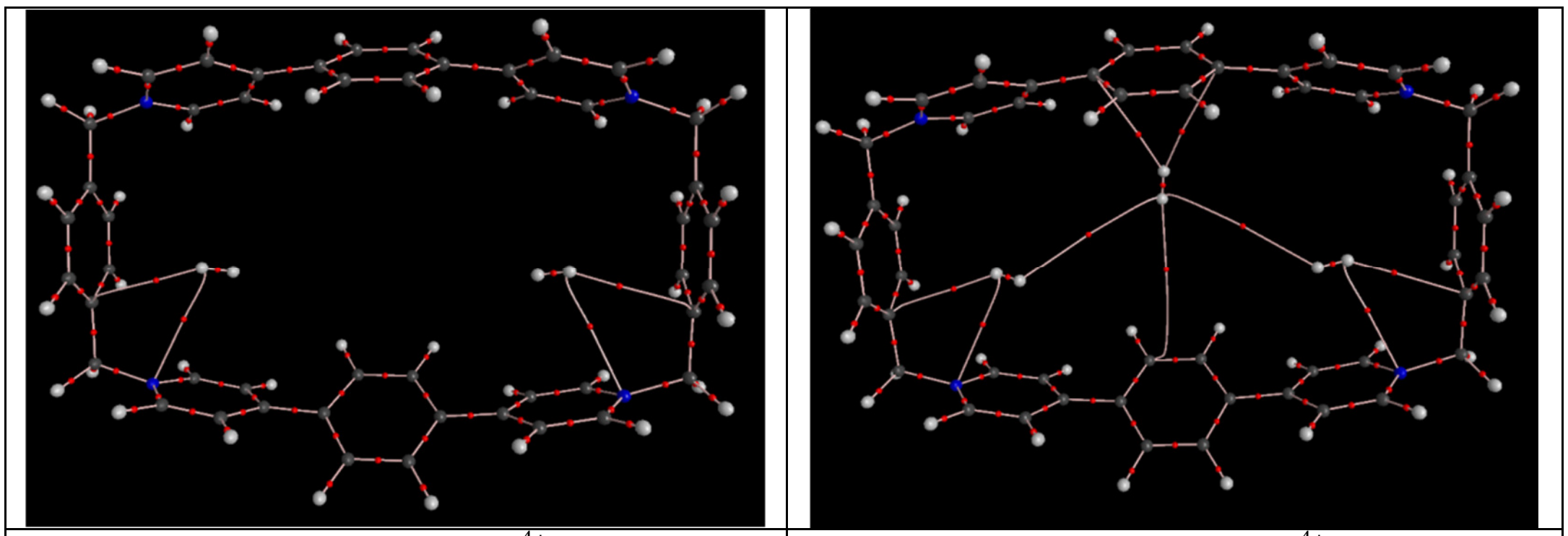

$(2H_2)_{endo}@ExBox^{4+}$ | $(3H_2)_{endo}@ExBox^{4+}$

Figure S1:Molecular graph showing bond paths, bond critical points(BCP, red dots) of $nH_2@ExBox^{4+}$

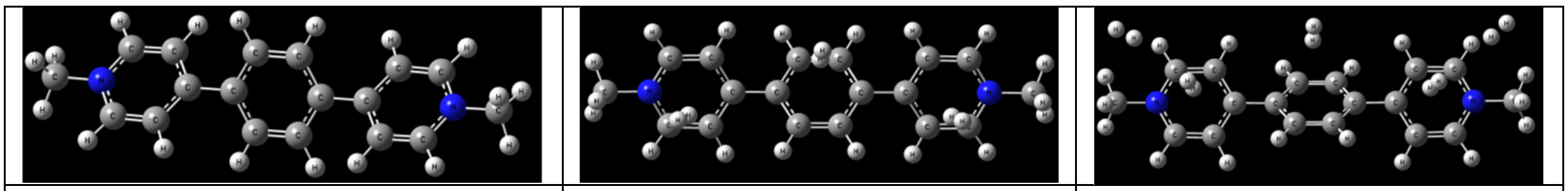

(a) | (b) | (c)

Figure S2: Structure of model and $nH_2$@model (n=3,5)

| | |
|---|---|
| 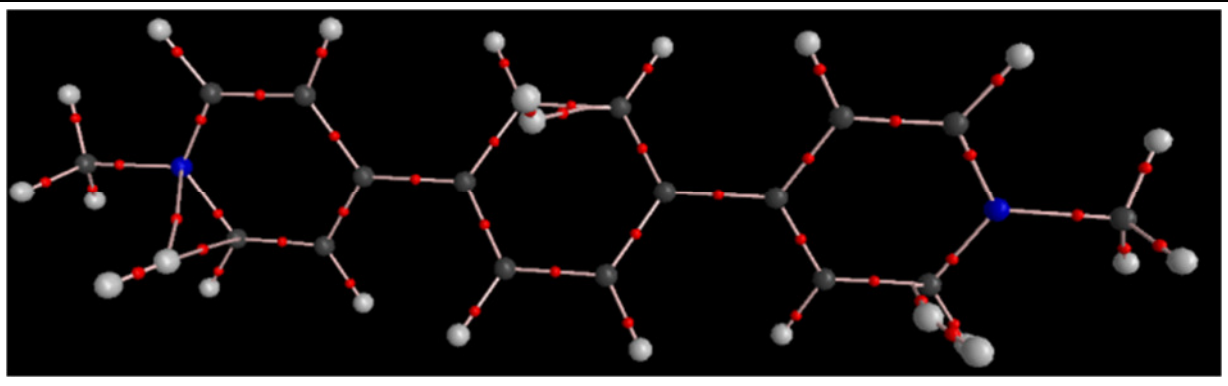 | 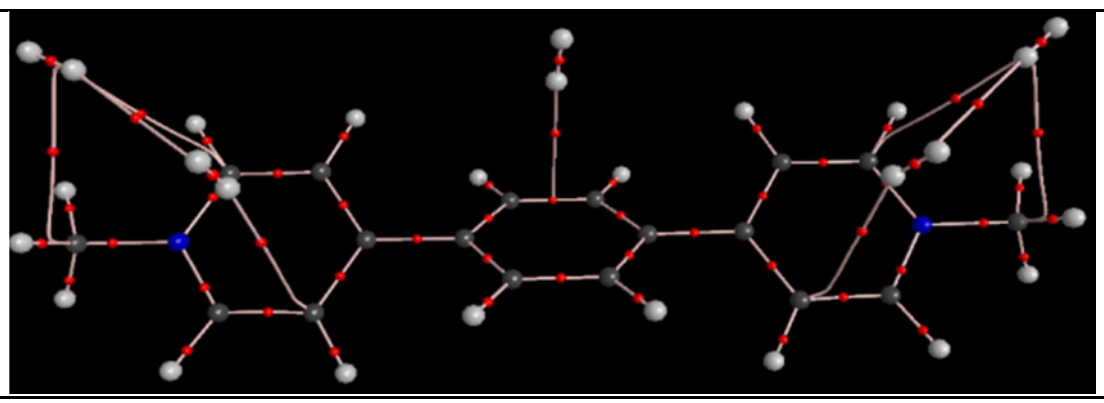 |
| | |

Figure S3:Molecular graph showing bond paths, bond critical points(BCP, red dots) of $nH_2$@model (n=3,5)

| | |
|---|---|
| 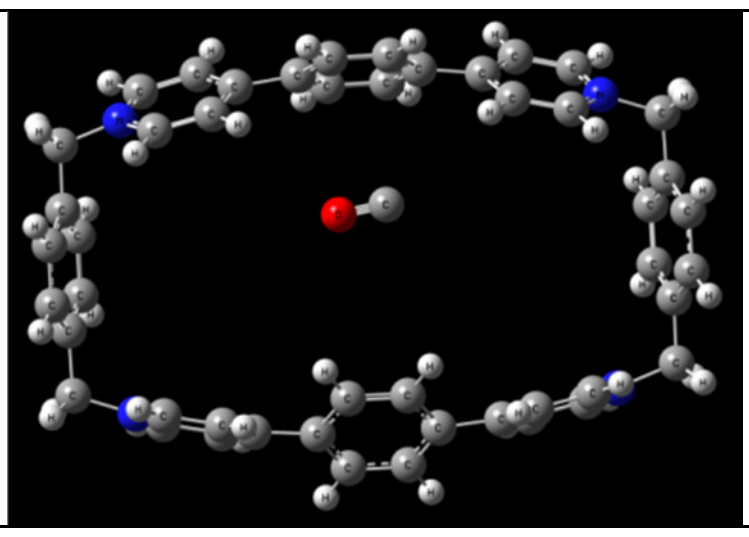 | 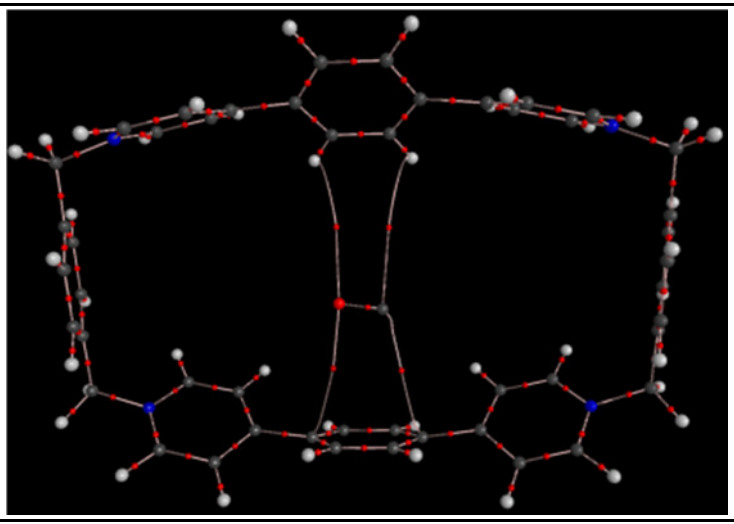 |
| Optimized geometry | Molecular graph showing BCP and bond path |
| $(CO)_{endo}$@$ExBox^{4+}$ | |
| 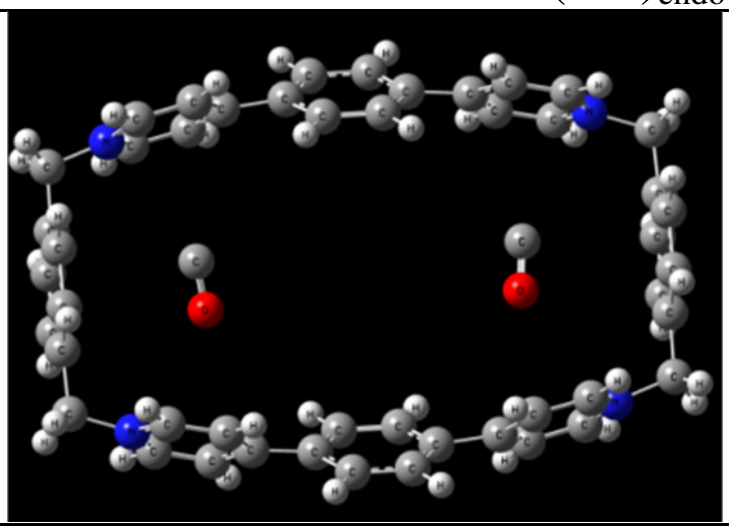 | 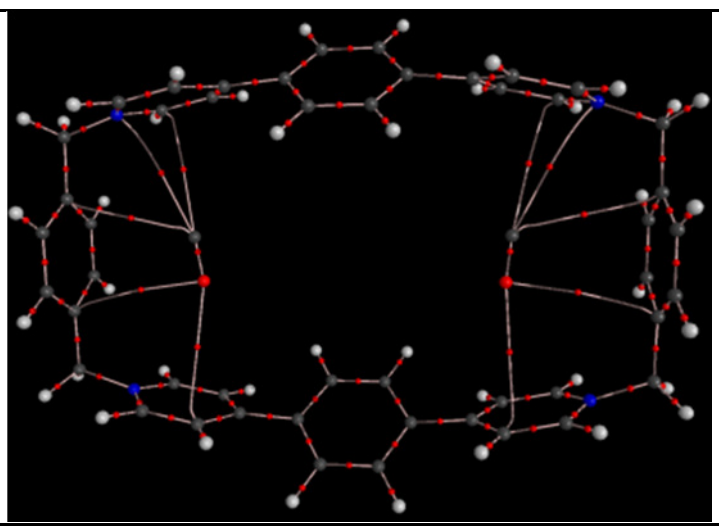 |
| Optimized geometry | Molecular graph showing BCP and bond path |
| $(2CO)_{endo}$@$ExBox^{4+}$ | |
| 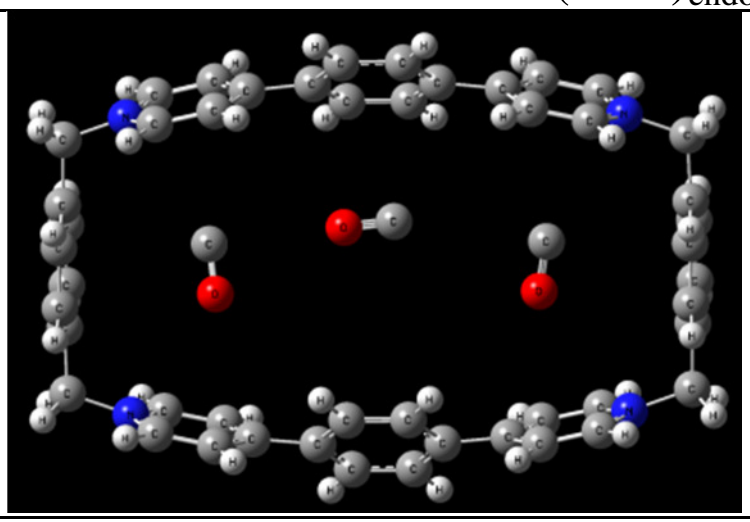 | 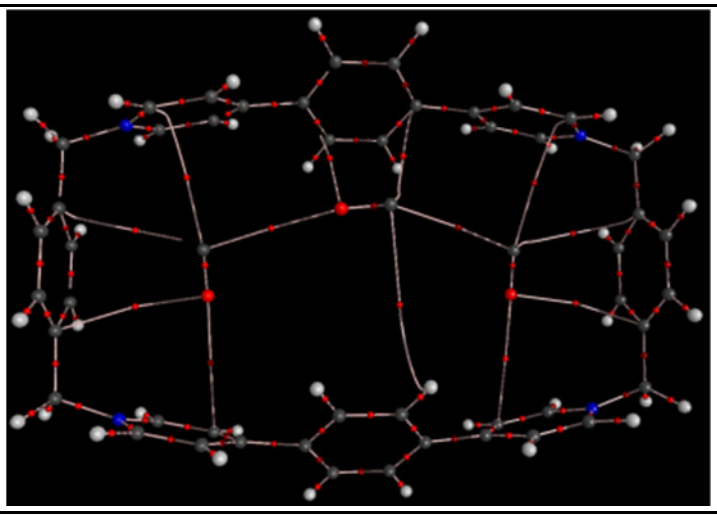 |
| Optimized geometry | Molecular graph showing BCP and bond path |
| $(3CO)_{endo}$@$ExBox^{4+}$ | |

Figure S4:Optimized geometry and molecular graph showing bond paths, bond critical points(BCP, red dots) of $(nCO)_{endo}$@$ExBox^{4+}$; n=1,2,3

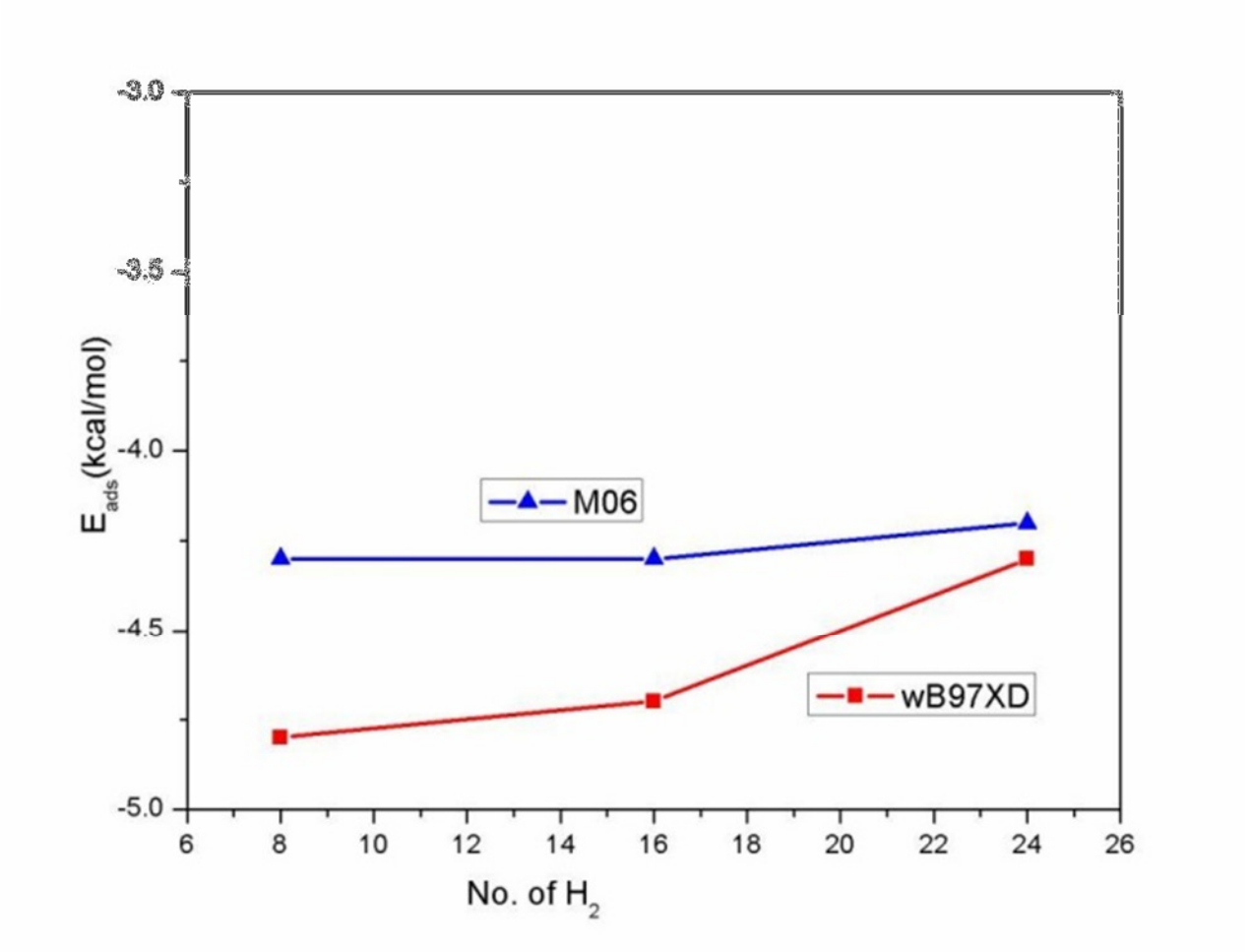

Figure S5: The plot of adsorption energy as a function of number of hydrogen adsorbed calculated at 6-311G(d,p) basis set.

<table>
<tr>
<td>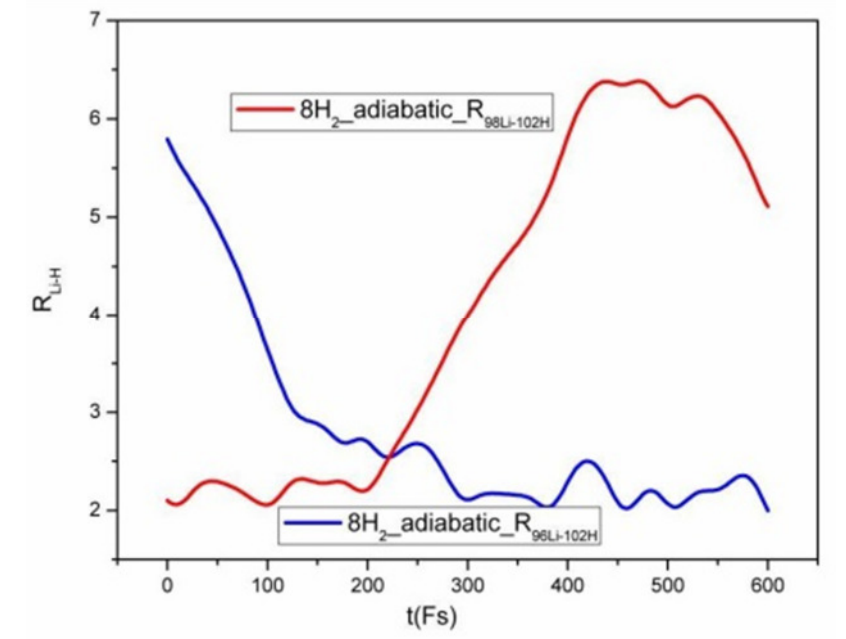
</td>
<td>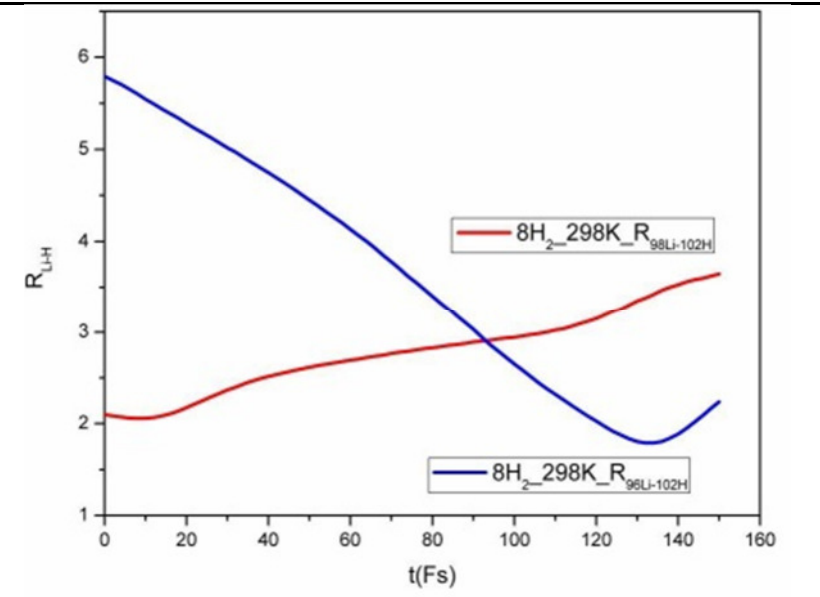
</td>
<td>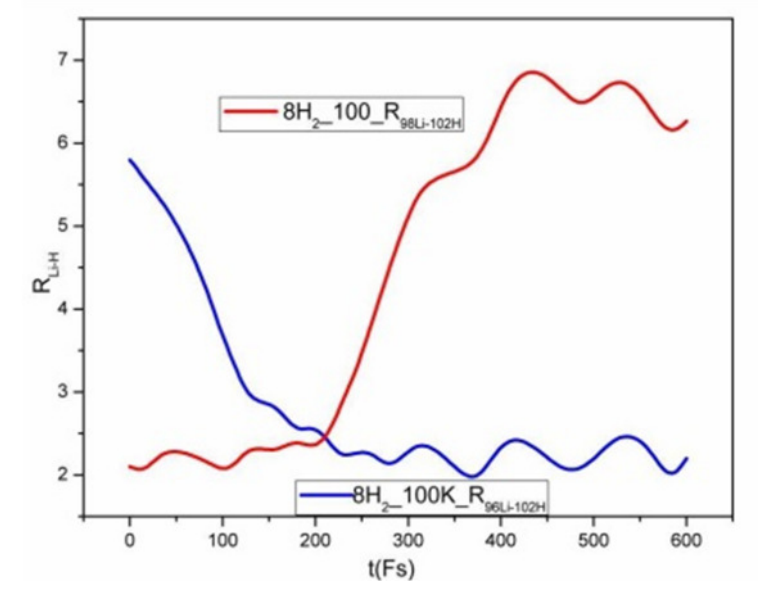
</td>
</tr>
<tr>
<td>Adiabatic method, Li-H bond length</td>
<td>298K simulation,  Li-H bond length</td>
<td>100K simulation, Li-H bond length</td>
</tr>
<tr>
<td>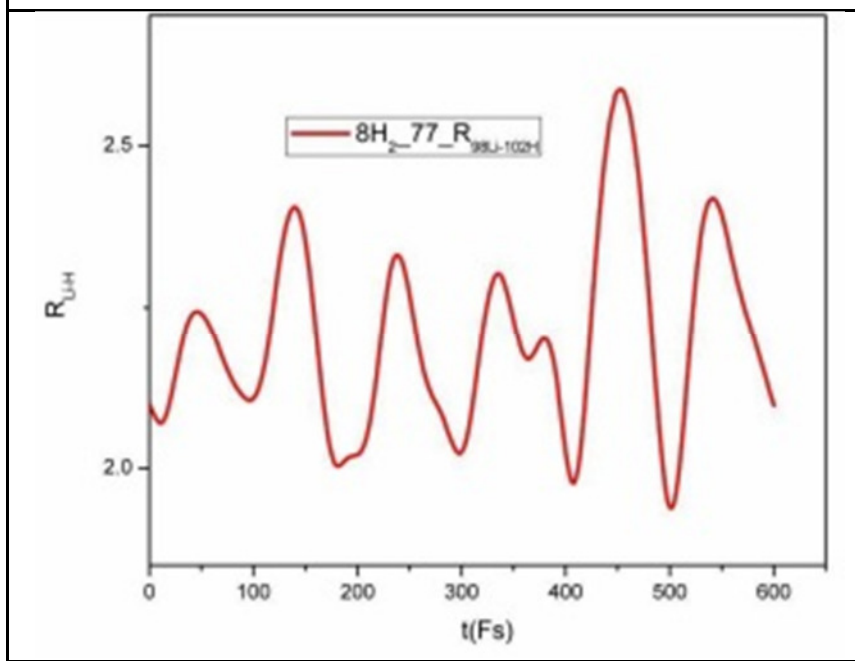
</td>
<td>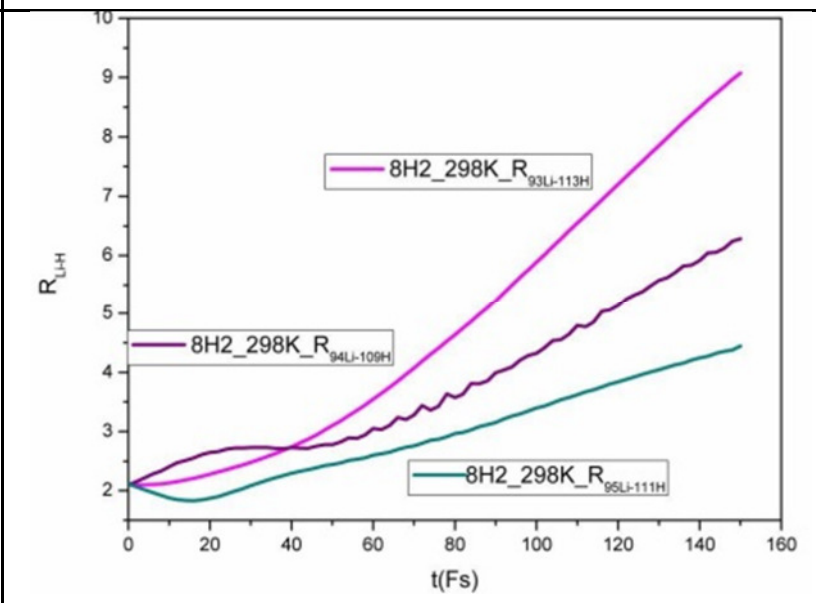
</td>
<td>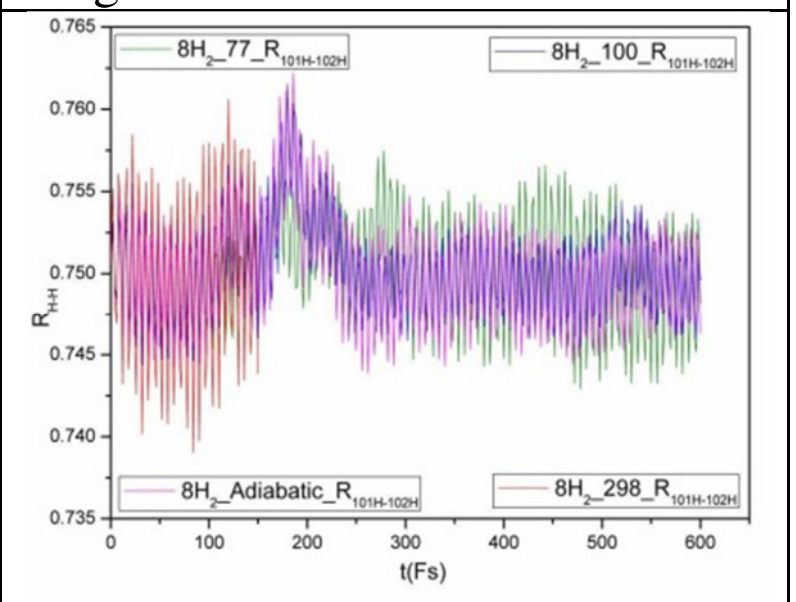
</td>
</tr>
<tr>
<td>77K simulation, Li-H bondlength</td>
<td>Li-H bond length  at 298K</td>
<td>H-H bod length at different temperatures,</td>
</tr>
<tr>
<td></td>
<td>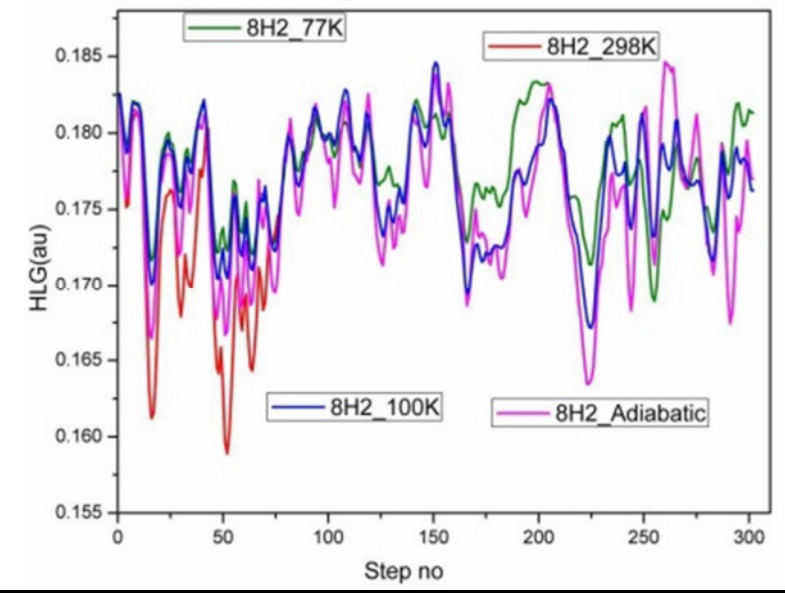
</td>
<td></td>
</tr>
<tr>
<td></td>
<td>Time evolution of HLG</td>
<td></td>
</tr>
</table>

Figure S6: Trajectory plot of the distance as a function of time between and time evolution of HLG for $8H_2$@ $Li_8ExBox^{4+}$.

<table>
<tr>
<td>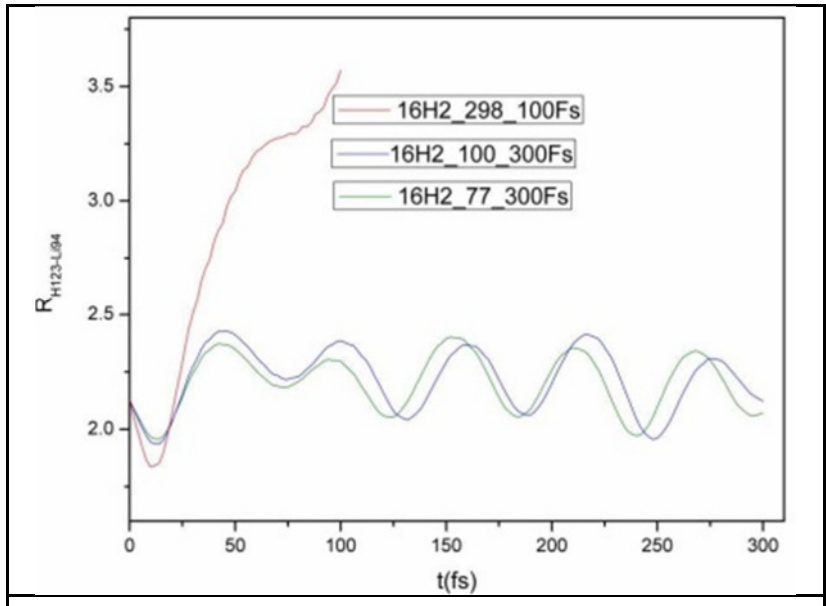
</td>
<td>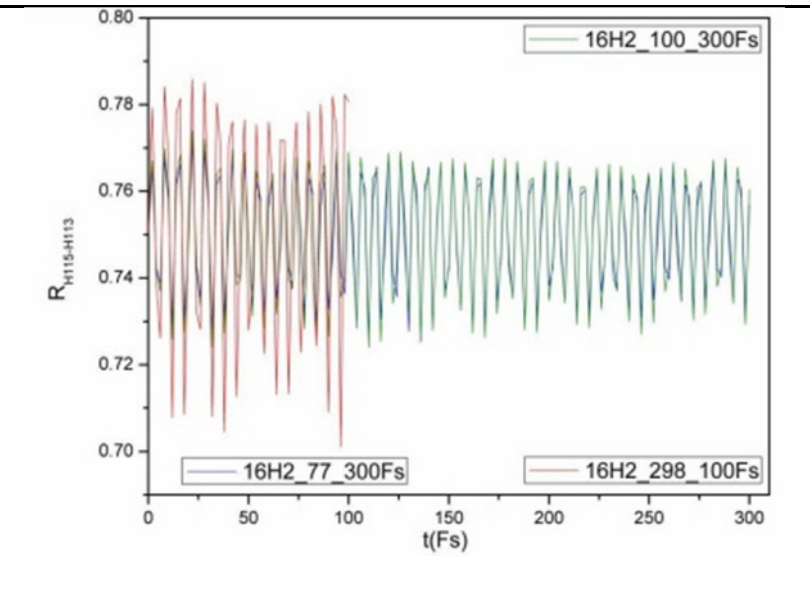
</td>
<td>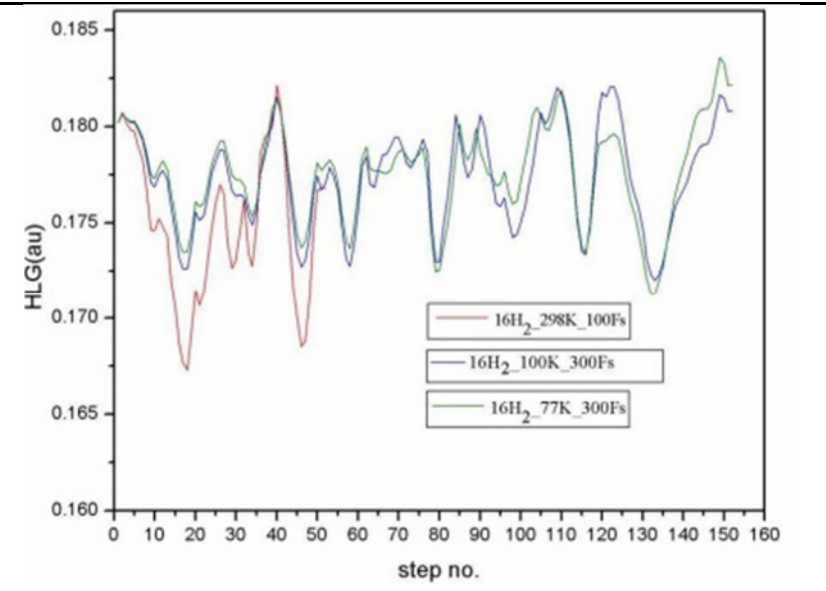
</td>
</tr>
<tr>
<td>Li-H bond length at different temperatures</td>
<td>H-H bond length at different temperatures</td>
<td>HLG variation</td>
</tr>
</table>

Figure S7:Trajectory plot of the distance as a function of time between and time evolution of HLG for $16H_2$@ $Li_8ExBox^{4+}$.

<table>
<tr>
<td>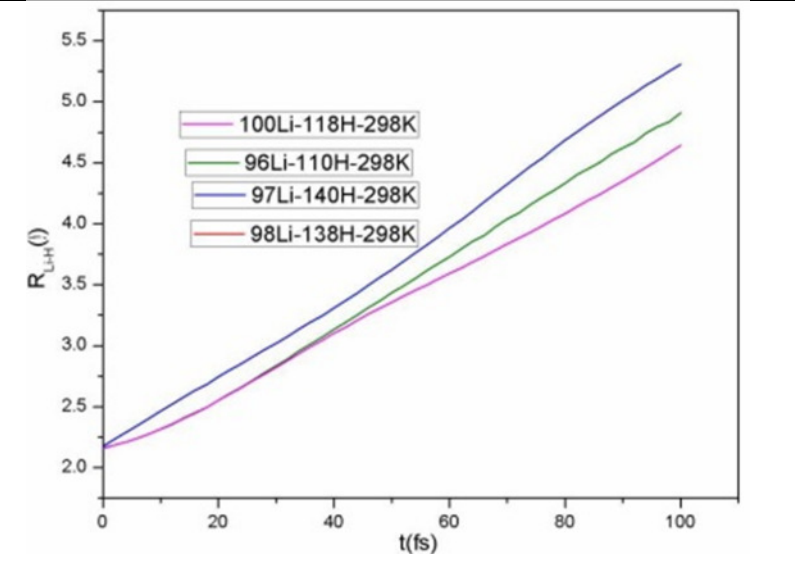
</td>
<td>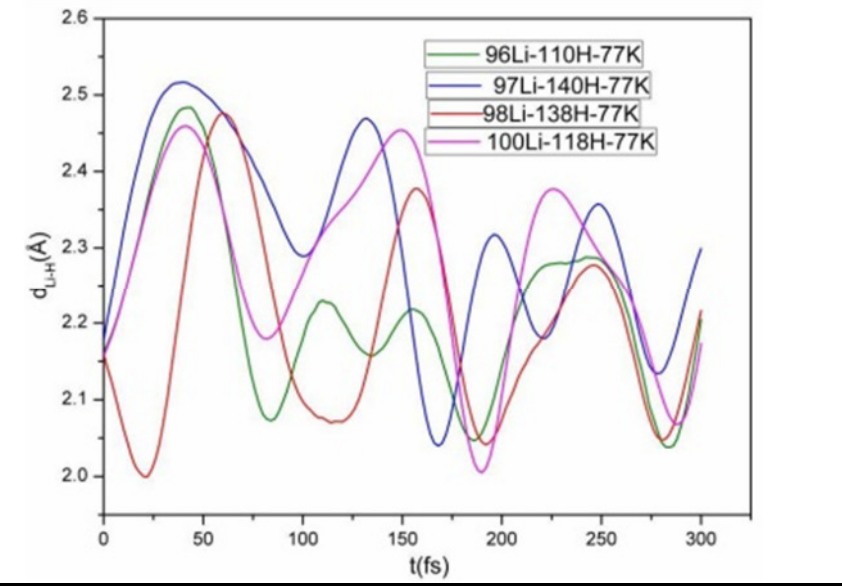
</td>
<td>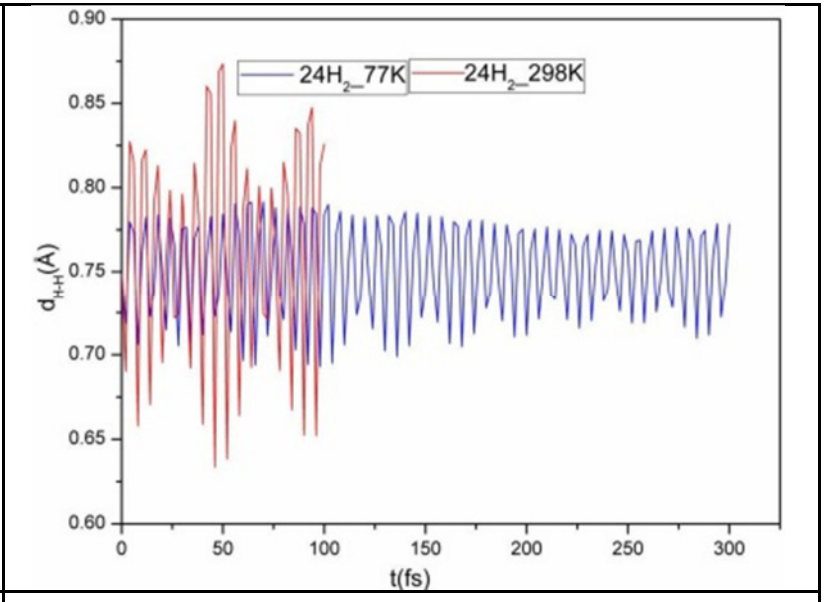
</td>
</tr>
<tr>
<td>Li-H bond length during 298K simulation</td>
<td>Li-H bond length during 77K simulation</td>
<td>H-H bond length during 298K and 77K simulation</td>
</tr>
<tr>
<td></td>
<td>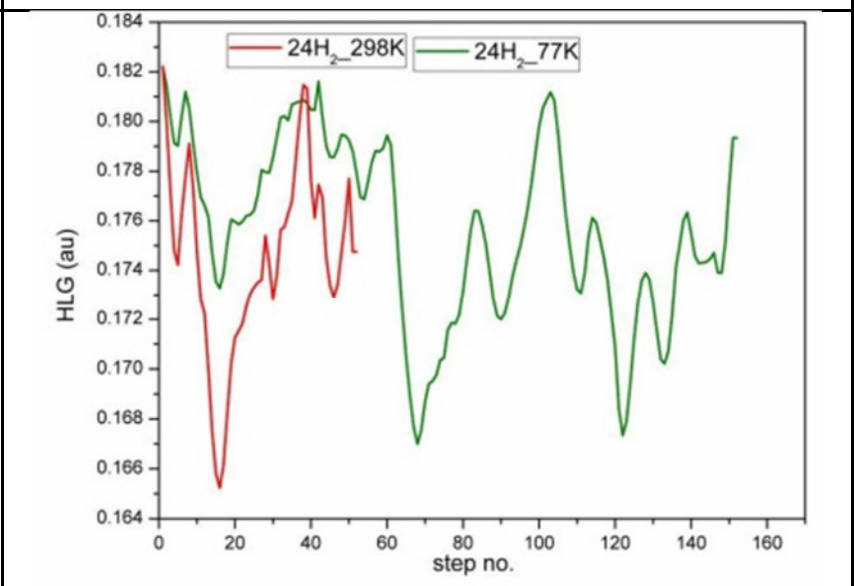
</td>
<td></td>
</tr>
<tr>
<td></td>
<td>HLG variation</td>
<td></td>
</tr>
</table>

Figure S8:Trajectory plot of the distance as a function of time between and time evolution of HLG for $24H_2$@ $Li_8ExBox^{4+}$.

<table>
<tr>
<td>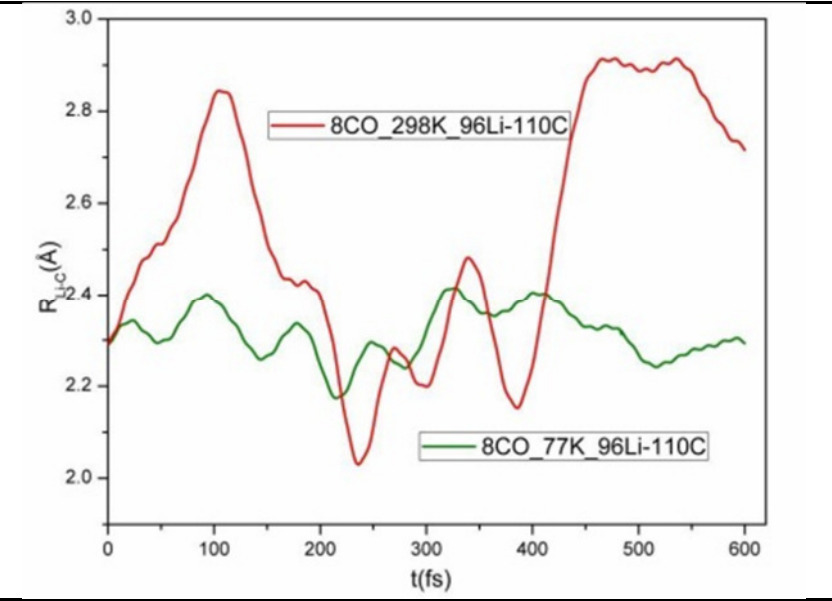<br></td>
<td>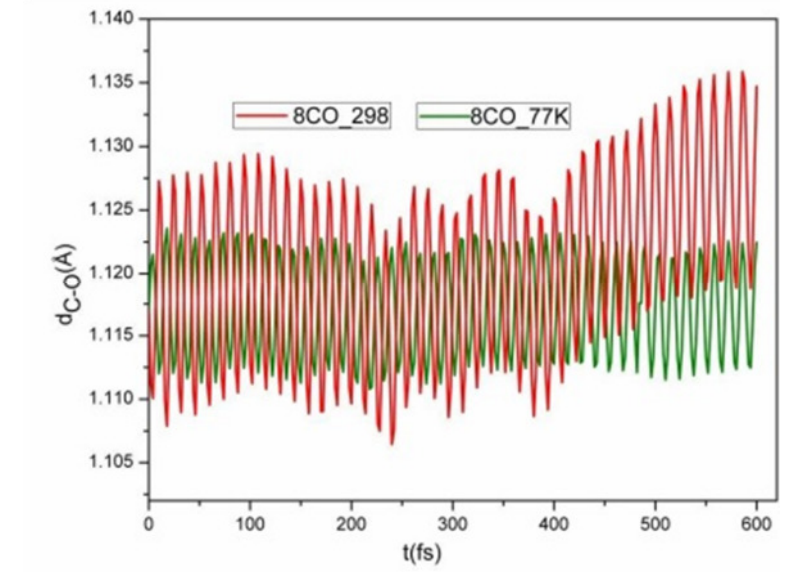<br></td>
<td>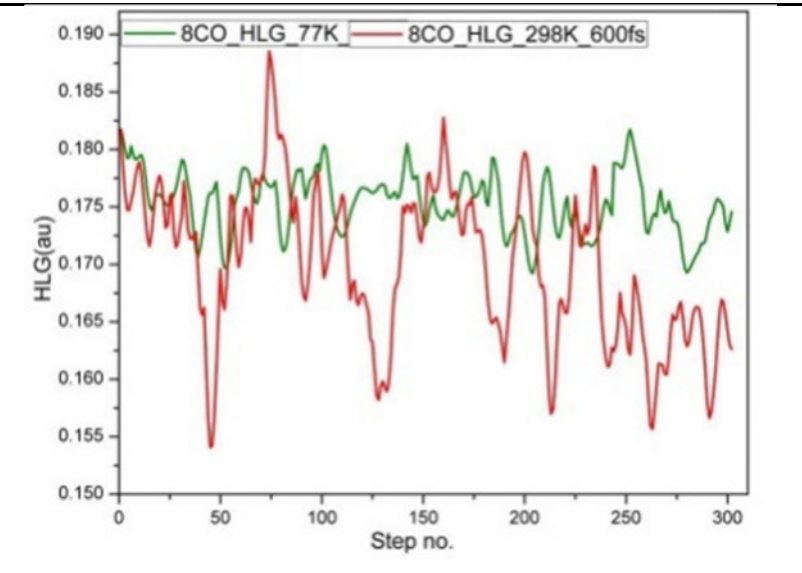<br></td>
</tr>
<tr>
<td>Li-C bond length variation at T=298 K,77 K</td>
<td>C-O bond length variation at T=298 K,77 K</td>
<td>HLG variation T=298 K,77 K</td>
</tr>
</table>

Figure S9: Trajectory plot of the distance as a function of time between and time evolution of HLG for 8CO@ $Li_8ExBox^{4+}$.

| | |
|---|---|
| 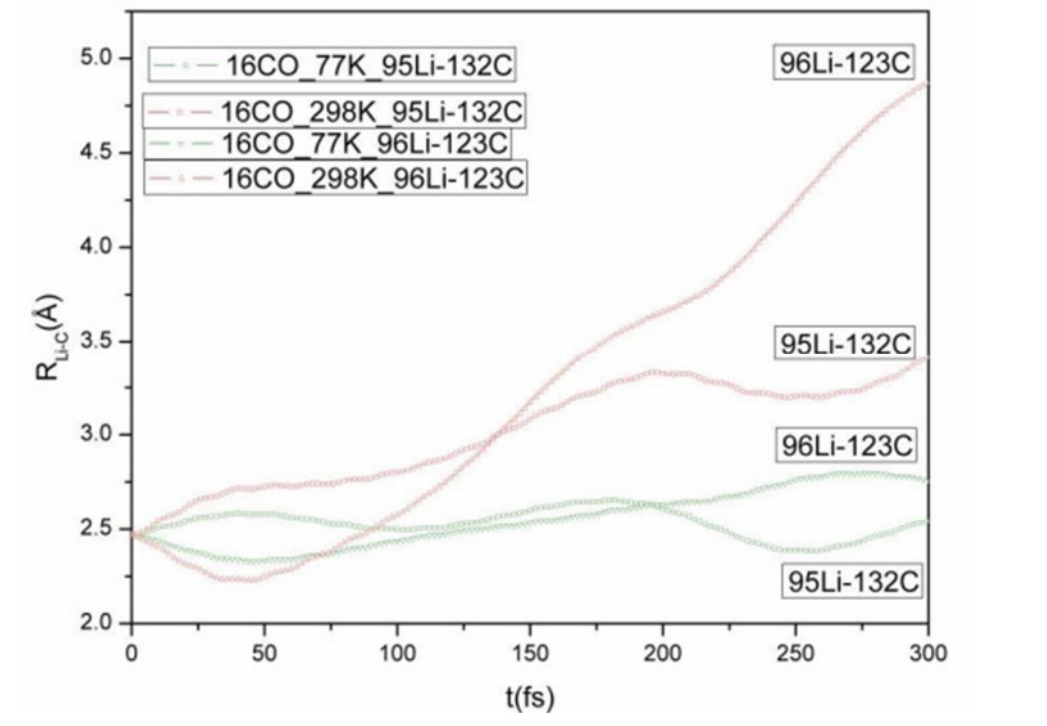  | 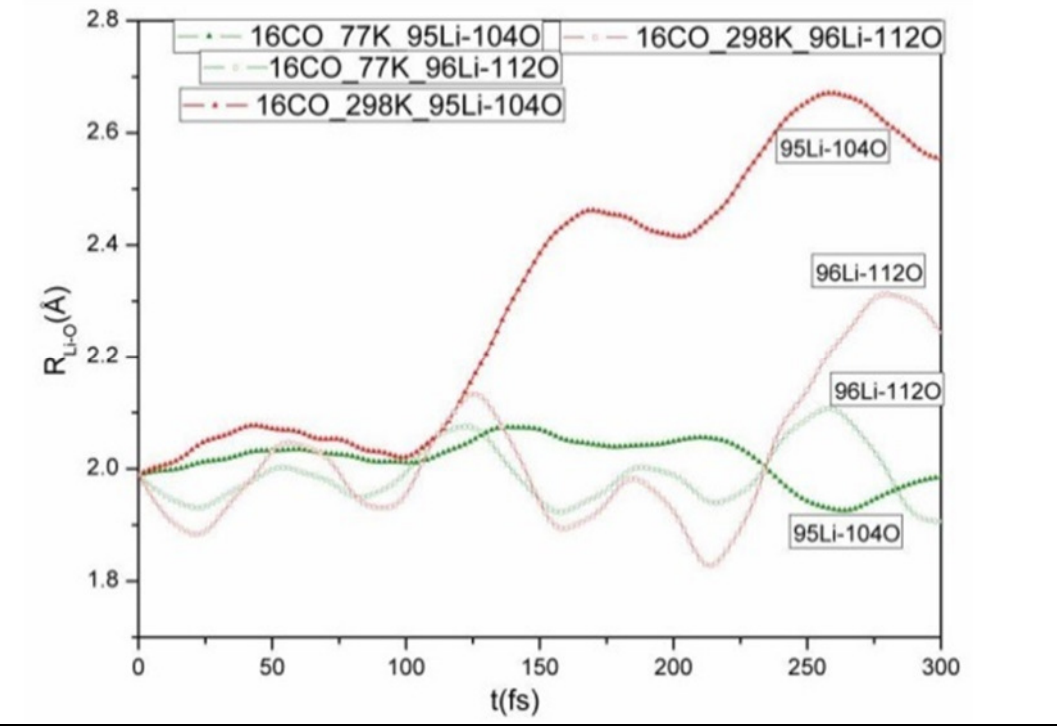  |
| Li-C bond length variation at T=298 K,77 K | Li-O bond length variation at T=298 K,77 K |
| 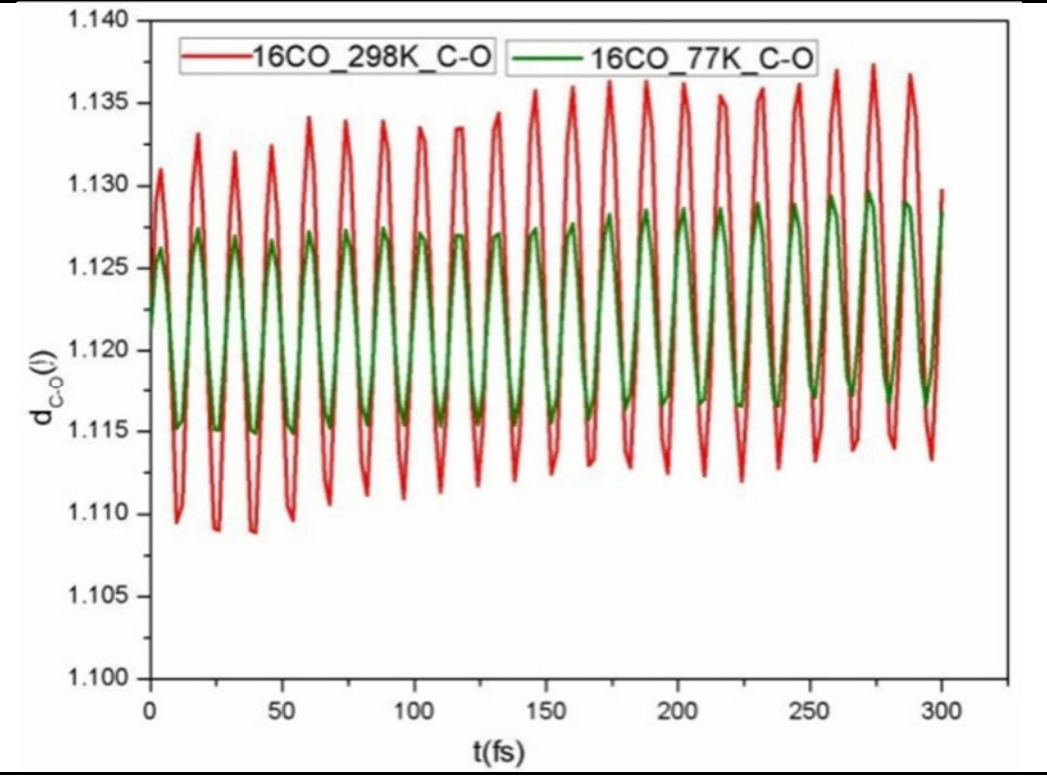  | 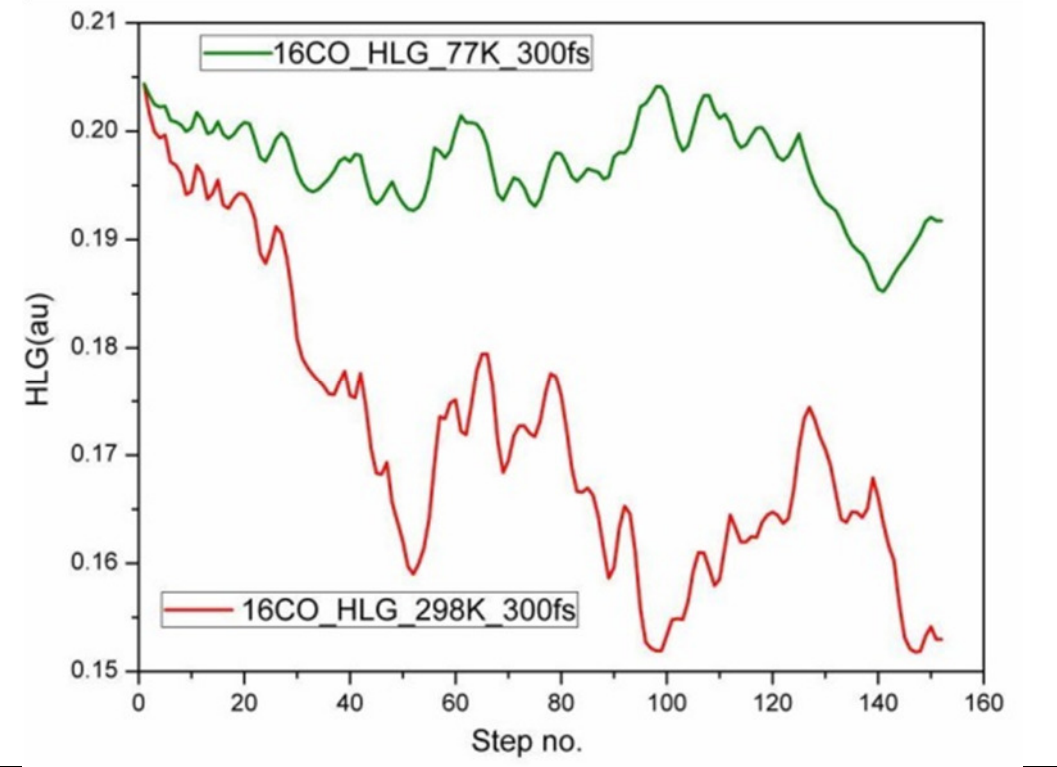  |
| C-O bond length variation at T=298 K,77 K | HLG variation T=298 K,77 K |

Figure S10: Trajectory plot of the distance as a function of time between and time evolution of HLG for 16CO@ $Li_8ExBox^{4+}$.

| | |
|---|---|
| 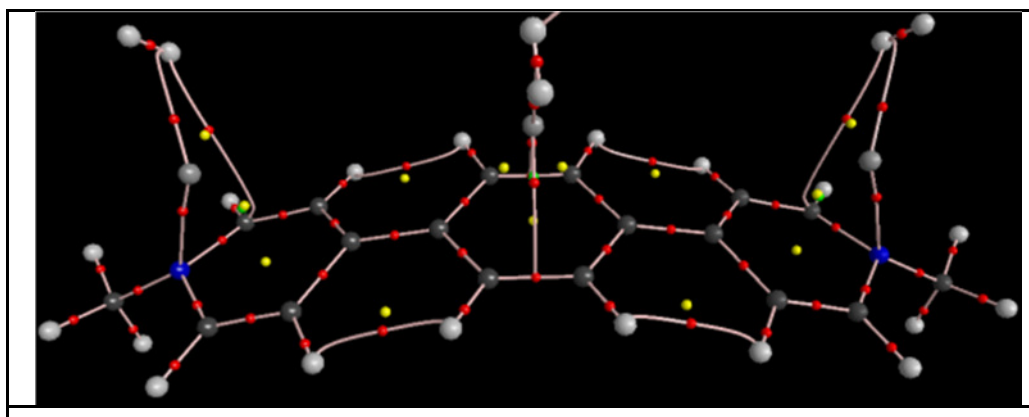 | 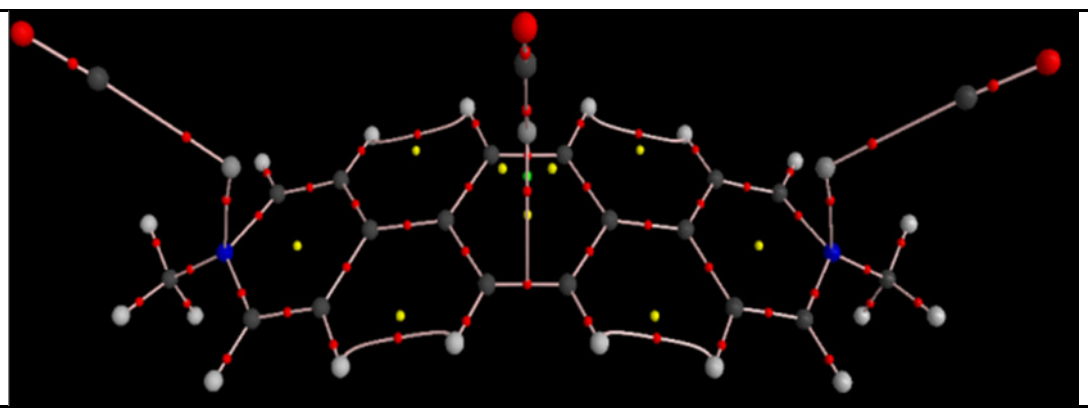 |
| $3H_2$@model | 3CO@model |

Figure S11: Molecular graph showing bond paths, bond critical points(BCP, red dots), ring critical points(RCP, yellow dots) and cage critical points(CCP, green dots).